\documentclass[preprint
 reprint,
 amsmath,amssymb,
 aps,
]{revtex4-2}

\usepackage{graphicx}% Include figure files
\usepackage{dcolumn}% Align table columns on decimal point
\usepackage{bm}% bold math
\usepackage{xcolor}

\newcommand{\myref}[1]{%
  \Roman{section}.\Alph{subsection}.\arabic{subsubsection}%
}

\begin{document}

\preprint{APS/123-QED}

\title{Macromolecular tribology at flowing solid/liquid interfaces}
%\title{Macromolecular tribology at solid/liquid interfaces}% Force line breaks with \\
%\thanks{A footnote to the article title}%

\author{Malo Velay}
% \altaffiliation[Also at ]{Physics Department, XYZ University.}%Lines break automatically or can be forced with \\
\author{Jean Comtet}%
 \email{jean.comtet@espci.fr}
\affiliation{%
Soft Matter Sciences and Engineering, ESPCI Paris, PSL University, CNRS, Sorbonne Université, 75005 Paris, France
}%

% Authors' institution and/or address\\
% This line break forced with \textbackslash\textbackslash

\date{\today}% It is always \today, today,
             %  but any date may be explicitly specified

\begin{abstract}

Molecular-scale interactions between solvated macromolecules and solid surfaces govern a large number of processes, from biology to engineering. Yet, despite extensive characterization at the macroscopic level, our molecular understanding of polymer/surface interactions remains limited, particularly under out-of-equilibrium conditions. Here, we combine wide-field single-molecule microscopy with microfluidic transport to directly track the nanoscale dynamics of individual fluorescently tagged macromolecular PEG adsorbates, and investigate their subtle couplings with interfacial hydrodynamic flows. At equilibrium, we evidence marked surface dependence, with macromolecular dynamics switching from  heterogeneous non-Brownian diffusion on hydrophilic glass to bidimensional Brownian-like transport in an interfacial physisorbed state on hydrophobic self-assembled monolayers. While for hydrophilic glass, the effect of the flow is restricted to an advective contribution during solvent-mediated flights, we uncover for the hydrophobic surfaces a peculiar regime of mixed macromolecular friction, whereby the adsorbed chain rubs on the solid wall while being continuously dragged by the near-surface hydrodynamic flow through interfacial slippage. Through joint analysis of equilibrium and out-of-equilibrium transport, we finely disentangle these molecular level frictional interactions with both the solid surface and the interfacial liquid. Beyond population-averaged dynamics, we further unveil a broad distribution of friction coefficients associated to individual chains, which we attribute conformational heterogeneities with sluggish reorganization timescale. By enabling direct observations of molecular-scale interfacial dynamics, our approach provides a novel molecular picture of macromolecular friction and adsorbate/surface interactions at flowing solid/liquid interfaces, and should guide molecular modeling of the macroscopic response of interfacial soft matter.

\end{abstract}

%\keywords{Suggested keywords}%Use showkeys class option if keyword
                              %display desired
\maketitle
\section{Introduction}

%The force-driven response of solvated macromolecules adsorbed to solid surfaces is key to numerous processes from biology to engineering. During transport in porous media \cite{cuenca2013submicron, park2019depletion}, membrane filtration, or (bio)sensing assays \cite{squires2008making}, macromolecular adsorbates of various nature interact with solid surfaces while being jointly submitted to hydrodynamic forces. Interfacial macromolecular interactions can also strongly affect emergent mechanical material response. Bio-lubricating macromolecules present in the eye or cartilage \cite{lin2022hydration} and synthetic hydrogels alike \cite{gong2006friction,ciapa2024friction,kim2016soft} possess remarkable frictional properties, attributed to peculiar interfacially-driven macromolecular interactions. Polymeric adsorption to solid surface can also provide enhanced dissipative pathways in hybrid nanocomposite gels \cite{rose2014nanoparticle} and allow for tunable adhesion between wet and soft materials \cite{rose2013time}.

The mechanically driven response of solvated macromolecules adsorbed onto solid surfaces underpins a broad spectrum of processes in both biology and engineering. Transport in porous media and membrane filtration \cite{cuenca2013submicron,park2019depletion,grzelka2021slip}, (bio)sensing assays or nanopore translocation \cite{squires2008making,wanunu2008dna}, blood clotting  \cite{sing2013willlebrand,schneider2007shear} or bacterial adhesion \cite{thomas2008biophysics} are all examples were macromolecular adsorbates—whether natural or synthetic—interact with solid surfaces while simultaneously experiencing hydrodynamic or electrostatic forces. Interfacially active macromolecules can also profoundly influence the mechanical properties of materials and interfaces. Biolubricating macromolecules found in ocular or cartilage systems \cite{lin2022hydration}, alongside synthetic hydrogels \cite{gong2006friction,ciapa2024friction,kim2016soft} exhibit exceptional frictional behaviors,  arising from the distinctive property of the macromolecular/solid couple. Polymer adsorption to solid surfaces can also be harvested for tunable adhesion between wet and soft materials \cite{rose2013time, biggs1995steric} or enhanced energy dissipation pathways in hybrid nanocomposite gels \cite{rose2014nanoparticle}. Additionally, surface-active macromolecules lead to distinctive lubricating properties \cite{zhu2002apparent}, which are exploited in diverse engineering applications—from the flow of colloidal or particulate suspensions \cite{chang1991effect} to hair conditioning \cite{adroher2023effect}. % also present as stabilizers from cement suspensions to hair conditionners.

%In tribology, the peculiar frictional properties of gels has been attributed to elastically-induced desorption of interfacially adsorbed macromolecules \cite{gong2006friction,ciapa2024friction}. Such adsorbed 
%or in ultra-low friction during boundary lubrication in biological settings such as cartilage \cite{lin2022hydration}.

These interfacial processes are typically probed at the ensemble level and over spatially extended areas, hiding the underlying complexity associated with the force-driven response of individual interfacial chains at the molecular scale. 
This mechanically driven chain motion is intrinsically complex, potentially involving interfacial sliding, conformational change, force-induced desorption or stick-slip dynamics. However, despite its major impact on the large-scale properties of the interface, this molecular-scale response remains obscure, being largely inaccessible to direct observation.

Regarding the question of surface-interacting polymers submitted to a hydrodynamic flow -- the specific focus of this  study -- early experimental 
investigations of macroscopic adsorption density yielded ambiguous results, with contrasting effects of the flow, ranging from force-induced desorption \cite{chang1991effect, lee1985adsorption, soga1998flow} to the absence of flow-induced response \cite{chin1991adsorption} as well as conflicting molecular-weight dependence \cite{lee1985adsorption,soga1998flow}. Recent molecular dynamics studies further revealed contrasting flow effects, including both chain flattening and lift~\cite{serr2010single}, a flow response strongly dependent on the adsorbed conformation~\cite{dutta2015shear,dutta2013adsorption}, as well as potentially complex in-plane polymer motion associated to slipping or rolling \cite{radtke2014shear} - yet with no experimental confirmations.

%Recent molecular dynamics simulations  suggest a possible reconciliation of these observations taking into account the degree of adsorption of the chains on the surface. Obtaining further insights would require to be able to separate in plan and out of plane motion.

% Single-polymer adsorption in shear: Flattening vs. hydrodynamic lift and surface potential corrugation effects

%\cite{he2010statistics} polymer under shear.

% --> shear induced globule

%\cite{radtke2015shear} ---> adsorption with catch bond

%profound influence on the la prorperties

%Yet the underlying chain motion under mechanical forcing can be potentially

%-- and with major impact on the larger-scale properties of the interface. 

%\textit{The underlying interfacial chain motion under such hydrodynamic or elastic mechanical forcing is expected to exhibit complex dynamics, with chain response possibly characterized by interfacial sliding, internal elastic deformation, force-induced desorption or stick-slip dynamics.  Yet, these interfacial processes are generally probed at the ensemble level and over spatially extended areas, hiding the underlying complexity associated with the force-driven response of interfacial chain at the molecular scale. This fundamental molecular-scale response thus remain largely inaccessible to direct observation, despite its profound influence on the macroscopic properties \cite{schroeder2018single}. \textbf{For example in the context of flow-induced polymeric response.. Even the equilibrium situation of pure diffusion remains challenging to address ...}}

Going back to macromolecular-scale dynamics, novel experimental approaches are now able to provide insights into the behavior of polymers at solid interfaces at the single-molecule level. Under equilibrium conditions, Atomic Force Microscopy (AFM) in liquid environments has been employed to image both DNA biomolecules~\cite{pastre2005study,guthold1999direct} and synthetic polymers~\cite{kumaki2008peculiar,kumaki2006reptational} at interfaces, uncovering conformational dynamics during surface diffusion. Equilibrium diffusion studies have also greatly benefited from advances in single-molecule fluorescence microscopy. Direct imaging of long DNA molecules adsorbed onto fluid-like lipid membranes revealed Rouse-like scaling (linear with chain length) of the friction coefficient~\cite{maier1999conformation}. For flexible synthetic polymers on solid surfaces, the situation proved more complex. Early experiments using Fluorescence Correlation Spectroscopy (FCS) suggested peculiar super-linear scalings of friction with chain length~\cite{sukhishvili2002surface,sukhishvili2000diffusion}. However, despite its single molecule sensitivity, FCS probes intensity fluctuations at a single point, leading to challenges in disentangling in-plane surface transport from out-of-plane desorption and exchanges with the liquid. These conclusions were thus revisited through direct molecular-scale imaging via single-molecule fluorescence tracking, revealing rather anomalous non-Brownian diffusive dynamics at the interface, with polymer dynamics consisting of alternating periods of chain immobilization at the surface, followed by long-distance jumps through solvent-mediated transfer~\cite{wang2016polymer,wang2015scaling,wang2020non,yu2014revisiting}. While these results contrast sharply with the traditional view of interfacial polymer diffusion as a Brownian process driven by monomeric friction, their generality remains to be confirmed: more continuous diffusive dynamics were evidenced for synthetic chains at the interface of liquid PDMS oil and water \cite{wang2015scaling}, and iny recent observations of single-strand DNA mobility at 2D material/water interfaces~\cite{shin2025diffusion}.% - yet with little quantitative characterizations.

%desai2007modeling \cite{qian2007surface,yu2014revisiting}.

Macromolecular response under force has received significantly less attention in comparison, despite both fundamental (probing force-induced transport beyond linear response theory) and applied motivations (many of the examples detailed above involve driven macromolecular transport).
Shear induced stretching at the interface could be revealed by AFM imaging \cite{he2009shear} following spin-coating.
AFM force spectroscopy techniques were also used to drag single tip-anchored chains, evidencing distinct frictional regimes ranging from kinetic sliding~\cite{kuhner2006friction} to cooperative stick~\cite{balzer2013nanoscale}. These studies revealed molecular friction and adhesion to be governed by near-surface interactions, such as hydrophobic~\cite{horinek2008peptide} or hydrogen-bonding effects~\cite{erbas2012viscous}.
Finally, in the context of the wall slip of polymeric solutions, confocal fluorescence measurements were used to probe the near-surface dynamics of fluorescent DNA molecules \cite{boukany2010molecular,hemminger2017microscopic}. These studies revealed that adsorbed DNA chains could either stretch or desorb under flow, yet the limited spatial and temporal resolution of these experiments were unable to finely disentangle the dynamics of adsorbed and near-surface chains and reveal in-plane motion of single chains.

Here, we couple state-of-the-art single-molecule microscopy techniques with microfluidic transport, to directly track the nanoscale dynamics of single fluorescently tagged water-soluble PEG macromolecules at solid/liquid interfaces and reveal their subtle interfacial coupling with hydrodynamic flow. The use of wide-field Single Molecule Microscopy allows to probe in-plane dynamics of interfacially adsorbed chains with superior spatial and temporal resolution, and analyze in detail the statistical properties of these interfacial molecular walks. We reveal marked dependence of interfacial dynamics on properties of the solid surface. On hydrophilic glass, individual PEG chains undergo strongly non-brownian surface diffusion, alternating between an adsorbed state and long desorption-mediated jumps through the solvent. Interfacial flow affects transport only through an advective contribution during solvent-mediated flights, leading to a subconvective scaling of the temporal displacement. On hydrophobic surfaces obtained by grafted Self Assembled Monolayers, we uncover instead a bidimensional brownian-like transport with chains diffusing a physisorbed state at the interface. During out-of-equilibrium transport, we uncover a peculiar transport mode, whereby the adsorbed PEG chains rub on the solid wall while being continuously dragged by the near-surface hydrodynamic flow. Joint analysis of equilibrium and out-of-equilibrium transport allows us to finely disentangle the frictional molecular interactions of the chains with both the wall surface and the solvent. Beyond the population averaged behavior, we go down to the analysis of the transport behavior of individual chains. We unveil broad distribution in the solid and liquid friction coefficients, which we attribute to heterogeneity in interfacial chain conformation, associated with slow reorganization timescales. By allowing for direct observations of local-scale interfacial dynamics, our approach brings a new molecular perspective of macromolecular friction and adsorbate/surface interaction at flowing solid/liquid interfaces.

%  , allowing to probe their subtle couplings with interfacial hydrodynamic flows. At equilibrium, we uncover an heterogeneous and strongly non-Brownian surface diffusion for individual chains, which alternate between a low-mobility adsorbed state and long desorption-mediated jumps through the solvent. The symmetry-breaking effect of the flow leads to a skewed distribution of interfacial displacements, with an unexpected dependence on the nature of the surfaces. On sticky hydrophilic surfaces, the hydrodynamic flow does not affect the chain motion except for an advective effect during solvent-mediated flights. 

%\subsection{Remaining citations}
% viscous friction HB MD

\section{Results and Discussions}
\subsection{Experimental Set-up}

% GPC distribution : see Moulab/Super_Resolution/GPC
\begin{figure}[htb]
\centering
\includegraphics[width=180 mm]{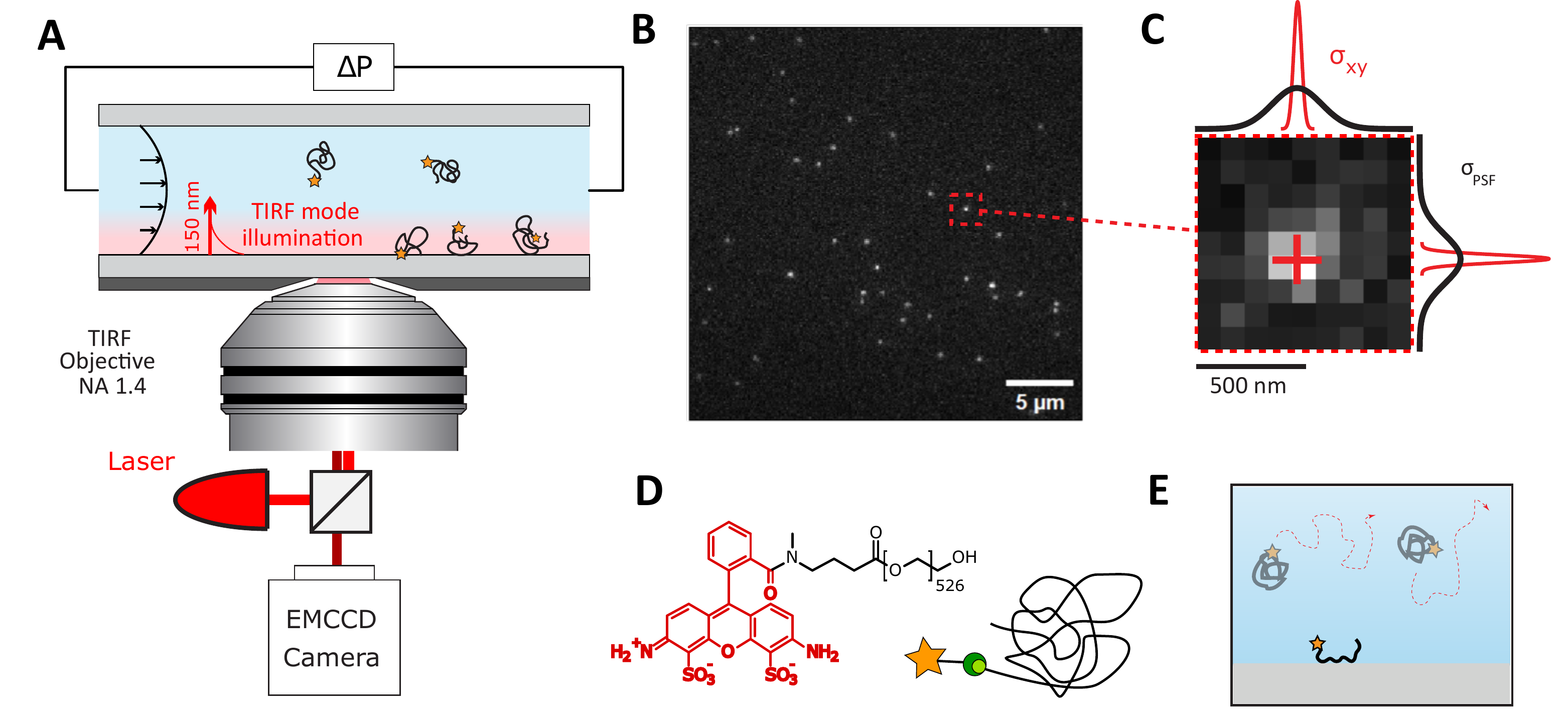}
\caption{\textbf{Monitoring single-polymer dynamics at interfaces under flow -} \textbf{(A)} Simplified schematic of the optical set-up coupled to a microfluidic device. The polymer solution is pushed through the microfluidic chip by a pressure controller. Illumination is in Total Internal Reflexion Fluorescence (TIRF) mode, leading to an exponentially decaying evanescent wave at the solid/liquid interface. \textbf{(B-C)} Typical image obtained: each diffraction spot corresponds to a macromolecule (B). Fitting the intensity distribution with a Gaussian Point Spread Function (C) allows to localize precisely the center of the fluorescent emitter, approximated as the position of the macromolecule. \textbf{(D)} Poly-(ethylene glycol), with $M_{\text{n}} = 20 \text{ kg.mol}^{-1}$ are functionalized by click-chemistry with a single fluorophore at the chain end. \textbf{(E)} Macromolecules  in the near-surface diffuse too fast to be localized at point emitters: fluorescent spots correspond only to  macromolecules adsorbed at the interface. }
\label{fig:Setup}
\end{figure}

A schematic of our experimental set-up is shown in Figure 1. Briefly, a single-molecule optical microscope in Total Internal Reflexion mode (TIRF) is used to image fluorescently tagged poly(ethylene) glycol (PEG) macromolecules interacting with the solid/water interface. Illumination in TIRF  allows us to probe the interfacial dynamic with high resolution by exciting the fluorescence over a thin layer above the surface of the glass coverslip with an evanescent wave (Fig. \ref{fig:Setup}A). A microfluidic chip coupled with a pressure controller imposes  a solvent flow at the interface, driving surface-adsorbed macromolecules  under out-of-equilibrium conditions. PEG chains are chemically functionalized with a single small and water-soluble fluorophore at the chain end (Fig. \ref{fig:Setup}D)  enabling the tracking of each chain with minimal perturbation of their dynamics. When these tagged chains adsorb to the interface, they appear as single fluorescent spots limited by diffraction on the camera sensor, (Fig. \ref{fig:Setup}B), corresponding to single Point-Spread functions (Fig. \ref{fig:Setup}C). These objects can be fitted with a gaussian function using super-localization techniques \cite{ovesny2014thunderstorm}, allowing to monitor the  center of the diffraction spot (assimilated to the center of mass of the PEG macromolecules), with a precision of approximately 20 nm. We stress that as the typical distance traveled by chains through bulk diffusion during our exposure time of $\Delta t =12 \text{ ms}$ reaches micrometric distances, only chains significantly slowed down through the adsorbing interaction with the surface are to be localized as single fluorescent spots (Fig. \ref{fig:Setup}E). This temporal effect accordingly leads to an effective nanometric uncertainty on the axial position of the molecule (see Materials and Methods section \myref{sec:MM_chemistry} for polymer chemistry and grafting, and section \myref{sec:MM_setup} for details on the experimental set-up and localization uncertainty).

\subsection{Qualitative observations of macromolecular dynamics on hydrophilic and hydrophobic surfaces}

%\subsubsection{A qualitative description of trajectory dynamics on hydrophilic and hydrophobic surfaces}

\begin{figure}[htb]
\centering
\includegraphics[width=180 mm]{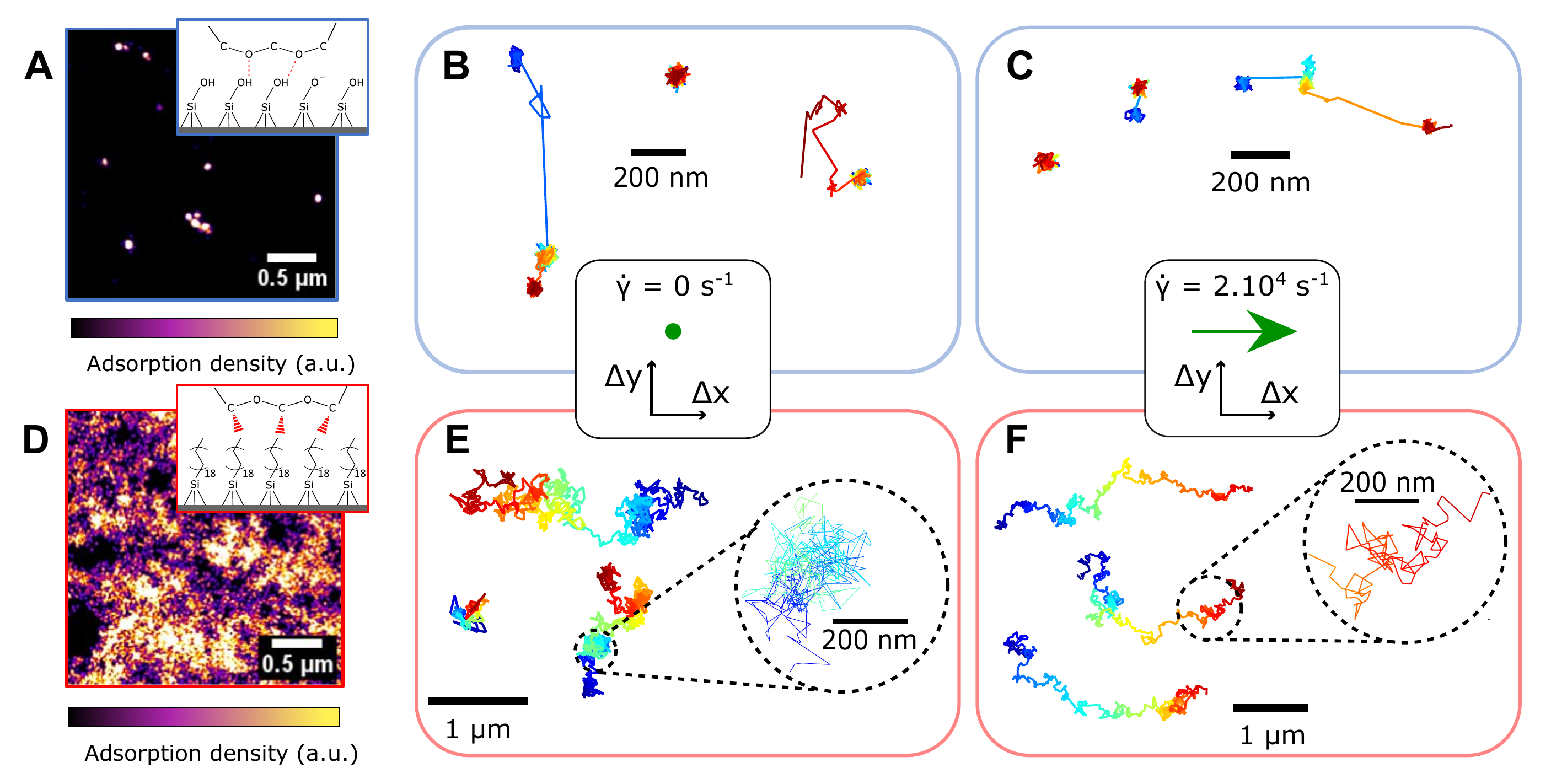}
\caption{\textbf{Representative trajectories on }\textbf{(A-C) }\textbf{hydrophilic bare glass surface and} \textbf{(D-F)} \textbf{hydrophobic silanized glass surface. }\textbf{(A, D)} Super-resolved map of the density of adsorption sites, with the adsorption density being color-coded. The surface chemistry of the corresponding interface is inserted on top of the image.  \textbf{(B, E)} Typical trajectories at equilibrium ($\dot \gamma=0 \text{ s}^{-1}$). \textbf{(C, F)} Typical trajectories under an out-of-equilibrium flow of solvent ($\dot \gamma=10^4$ s$^{-1}$, flow from left to right). Trajectory time is color-coded (from dark blue to red) with typical total times ranging from 2 to 3 s for B-C, 1 to 10s in E and about 5s in F.}
\label{fig:Traj}
\end{figure}
Temporal acquisitions can then be used to precisely track and quantify the associated interfacial macromolecular dynamic, which we report in Figure \ref{fig:Traj}. Two different classes of surfaces were considered to tune the molecular nature of the chain - surface interactions. Bare glass surfaces, washed with piranha solution, presents a hydrophilic surface covered with silanol groups, which are able to form hydrogen bonds with the PEG chains (inset Fig. \ref{fig:Traj}A). We considered in parallel silanized surface, which were rendered hydrophobic through grafting with a dense layer of alkyl chains, in which case the PEG chains can adsorb through hydrophobic interactions (inset Fig. \ref{fig:Traj}D). See Materials and Methods section \myref{sec:MM_surface} for details of surface preparation.
\newline
\paragraph*{\textbf{Hydrophilic glass.}}
We focus first on bare hydrophilic glass surfaces. Using super-resolution approaches, we first map the density and spatial positions of adsorption sites on the surface by summing the localizations from successive frames, as schematically represented in Fig.~\ref{fig:Traj}A. These reconstructed images highlight a spatially heterogeneous adsorption profile, with the presence of localized patches corresponding to high densities of localization events. We then turn to single-chain trajectories, which we obtain by correlating successive fluorescent events. Representative trajectories are shown in Figs. \ref{fig:Traj}B-C, respectively at equilibrium, i.e. $\dot \gamma=0$ s$^{-1}$ (Fig. \ref{fig:Traj}B) and under an applied shear rate of $\dot{\gamma} = 10^4$~s$^{-1}$ (Fig. \ref{fig:Traj}C). Raw movies are included in SI Video.S\ref{itm:SI_hydrophile_shear}. At equilibrium, the trajectories consist of localized adsorption events punctuated with long hopping displacements. This behavior aligns with previous reports on the interaction of PEGs with various surfaces, including fused silica~\cite{yu_single-molecule_2013} and hydrophobic surfaces obtained by grafting disordered tetramethylsilane monolayers~\cite{skaug_single-molecule_2014}. As shown in Fig.~\ref{fig:Traj}C, the applied shear rate has a clear effect on the trajectories, introducing a bias along the main flow direction. While the trajectories are visibly oriented, displacements in the transverse direction remain possible, and movement against the flow are not entirely prevented.

\paragraph*{\textbf{Silanized hydrophobic glass.}}
Turning to the silanized hydrophobic surfaces (Fig.~\ref{fig:Traj}E), we observe adsorbed chains to exhibit a fundamentally distinct dynamical behavior than on the previously considered hydrophilic surface, even under equilibrium conditions. This striking contrast between the two surfaces is clearly evident in the super-resolved image of Fig.~\ref{fig:Traj}D, which reveals an adsorption profile much more homogeneous and spatially continuous than in Fig.~\ref{fig:Traj}A. As shown by the trajectories reported in Fig.~\ref{fig:Traj}E, the transport dynamics appear qualitatively closer to classical Brownian-like motion, characterized by a more continuous distribution of jumping distances that form long and regular trajectories (see raw movies as SI Videos S\ref{itm:SI_hydrophobe_noshear} and S\ref{itm:SI_hydrophobe_shear}). Turning on the applied shear, we observe a continuous entrainment of the chain along the flow direction (Fig.~\ref{fig:Traj}F), characterized by what resembles again to a relatively continuous motion, with an associated drift speed of the order of $v_\text{drift} \approx 10$~nm$\cdot$s$^{-1}$. While it is tempting to compare this drift velocity to the interfacial fluid velocity expected on such surfaces—where grafted carbonaceous chains are expected to induce a finite slip length of $\lambda \approx 10-20$~nm—the corresponding value of fluid flow speed $v_\text{surface} = \lambda \dot{\gamma} \approx 1000$~nm$\cdot$s$^{-1}$ is up to three orders of magnitude larger than the drift velocity $v_\text{drift}$ observed here. This discrepancy suggests that molecular frictional interactions between the chain and the solid surface do play a dominant role in governing this interfacial sliding dynamics.

\subsection{Intermittent dynamics on hydrophilic interfaces}

To achieve a more refined and quantitative understanding of the modes of interfacial transport, we turn to the statistics of the displacement distributions. Individual trajectories are decomposed into successive steps, each representing elementary incremental displacements $\delta x = x(n+1)-x(n)$ and $\delta y = y(n+1)-y(n)$ respectively along \textit{(\textit{Ox}) }and transverse \textit{(Oy)} to the flow direction, with $n$ the frame number.  

Formally, these displacement distributions are characterized by the self part of the van Hove function \cite{hansen2013theory,van1954correlations}, which represents the probability that a molecule has moved a distance $\Delta x$ or $\Delta y$ along the $x$ or $y$ coordinate during time $\Delta t$, and can be expressed as \begin{equation*}
    G_s(\Delta x,\Delta t) = \frac{1}{N_\text{traj}} \left\langle\sum_{i=1}^{N_\text{traj}} \delta(\Delta x+x_i(t)-x_i(t+\Delta t)\right\rangle \text{ and }G_s(\Delta y,\Delta t) = \frac{1}{N_\text{traj}} \left\langle\sum_{i=1}^{N_\text{traj}} \delta(\Delta y+y_i(t)-y_i(t+\Delta t)\right\rangle
\end{equation*}
where the sum runs over all trajectories, $x_i(t)$ and $y_i(t)$ characterizes the displacements along the trajectories and $\langle.\rangle$ indicates ensemble averaging. These distributions can be accordingly evaluated at various lag times $\Delta t = n \tau_0$ with with $\tau_0 = 12$~ms as the experimental sampling time. Examining their statistical properties, allows us to  elucidate the precise mechanisms governing macromolecular dynamics at the interface.

\begin{figure}[htb]
\centering
 \includegraphics[width=180mm]{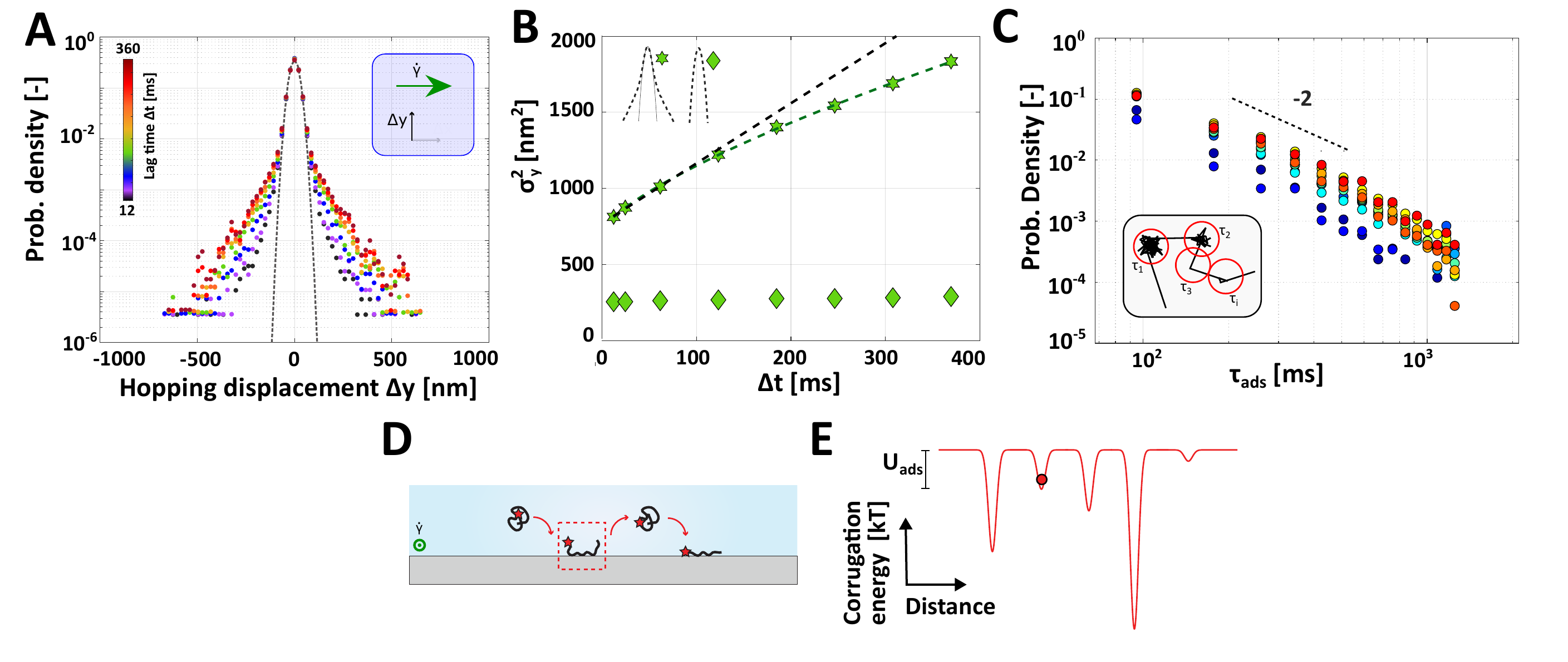}
\caption{\textbf{Transverse macromolecular dynamics at a sticky hydrophilic interface.} \textbf{(A)} Typical displacement distributions in the transverse direction on a hydrophilic surface for sampling times $\Delta t$ ranging from 12 to 360 ms. The interfacial shear rate is $\dot{\gamma} \approx 4 \times 10^{4}$~s$^{-1}$. The dashed black line corresponds to a gaussian fit to the central part of the distribution. \textbf{(B)} Evolution of the variance of the full distribution ($\sigma^2_\text{all}$, stars) and the central Gaussian component ($\sigma^2_\text{gauss}$, diamonds) at various sampling times $\Delta t$. The solid line represents the expected linear growth for Fickian diffusion, while the green dotted line indicates the sub-linear power-law evolution of $\sigma^2_\text{all}$ over time. \textbf{(C)} Distribution of local adsorption times at various shear rates, color-coded from blue ($0 \text{ s}^{-1}$) to orange ($2 \times 10^{4} \text{ s}^{-1}$). Inset shows schematic of the segmentation strategy to identify single adsorption times (See Material and Methods Section \myref{sec:MM_data} for details). \textbf{(D)} Local trajectory dynamics, highlighting the alternation between adsorption and hopping events. \textbf{(E)} Schematic of the broad, glassy-like adsorption energy landscape for adsorption.}
\label{fig:Phile1}
\end{figure}

\paragraph{\textbf{Heterogeneous and sub-diffusive transverse dynamics}}

We begin by examining Fig.~\ref{fig:Phile1}A, which shows the  distribution of displacements $\Delta y$ in the direction transverse to the flow on a bare glass surface. To elucidate the transport dynamics, these distributions are here plotted for increasing lag times ranging from 12 to 360 ms (from purple to red).

A prominent feature —consistent with previous reports~\cite{yu_single-molecule_2013,skaug_single-molecule_2014}— is the strongly non-Gaussian shape of the distributions, characterized by a narrow central peak and extended tails. As indicated by the dotted-line fit, the central peak is well-described by a Gaussian distribution with a width $\sigma = 16 \text{ nm}$, closely matching the typical localization uncertainty of about 20~nm and remaining independent of sampling time. This central mode corresponds to periods of immobilization, reflecting transient chain adsorption at the interface, as evident in the trajectories shown in Fig.~\ref{fig:Traj}B.

In contrast, the long-distance tails of the distribution broaden with increasing lag time $\Delta t$. These tails correspond accordingly to transient volumetric excursions of the macromolecules into the solution, followed by their readsorption at the interface. The time-dependent growth of these tails underscores the diffusive nature of this transport mode, illustrated in Fig.~\ref{fig:Phile1}D, where macromolecules alternate between adsorbed states and hopping events associated to 3D volumetric excursions.
To further quantify these distributions, we display in Fig.~\ref{fig:Phile1}B the variance of the total jumping distribution, $\sigma_\text{tot}^2$ (stars), alongside the central Gaussian component, $\sigma_\text{gauss}^2$ (diamonds), as functions of sampling time. Note that $\sigma_\text{tot}^2$ is quantitatively equivalent to the Mean-Squared Displacement defined as $\text{MSD}(\Delta t) = \langle [x(t+\Delta t) - x(t)]^2 \rangle$.
While $\sigma_\text{gauss}^2$ remains invariant with time, the ensemble variance, $\sigma_\text{tot}^2$, demonstrates time-dependent growth. Notably, this growth deviates from classical Fickian behavior ($\sigma_\text{tot}^2 \sim t$, black dashed line) and instead follows a sub-linear scaling law, $\sigma_\text{tot}^2 \approx t^\beta$, with here $\beta \approx 0.8<1$ (green dashed line). This sub-diffusive behavior arises from the interplay between the two transport modes depicted in Fig.~\ref{fig:Phile1}D, dominated in particular by the strongly adsorbing steps. From the initial linear fit (Fig.~\ref{fig:Phile1}B, black dashed line), we extract an equivalent diffusion coefficient of  $D = 4.5 \times 10^{-15} \text{ m}^2 \text{.s}^{-1}$ as $\sigma_\text{tot}^2=2Dt$, which is here five orders of magnitude smaller than the bulk diffusion coefficient $D_\text{bulk} \approx 4\cdot 10^{-11}$ m$^2$.s$^{-1}$ \cite{wang2015scaling,waggoner1995diffusion}, indicative of the extreme slowdown induced by surface adsorption.

To finely characterize the intermittency of the transport dynamics, we perform trajectory segmentation to extract the characteristic duration of each adsorbed period (see Materials and Methods III.E). The resulting distributions of adsorption times, shown in Fig.~\ref{fig:Phile1}C for increasing interfacial shear rates, exhibit a power-law behavior characterized by an exponent $\alpha \approx -2$.  This power-law scaling highlights the broadness of this distribution of adsorption times, contrasting with the exponential distribution expected for a single adsorption energy. As schematized in Fig.~\ref{fig:Phile1}E, the free energy profile can be accordingly described by a wide range of adsorption well depths, reminiscent of a glassy energy landscape. As previously suggested \cite{yu_single-molecule_2013,skaug_single-molecule_2014}, such broad distributions may arise from specific polymeric effects, associated in particular to the potentially broad range of conformations in the adsorbed state. Power-law exponents ranging from -1.6 \cite{yu_single-molecule_2013} to -2.5 \cite{skaug_single-molecule_2014,skaug2013intermittent} have been obtained on similar polymeric systems. Our scaling exponent $\alpha \approx -2$, falls within this range, with variations attributed tentatively to the combined effect of broad adsorption configurations and spatial surface heterogeneity  \cite{yu_single-molecule_2013}.
However, we note that similar power-law scalings in adsorption time at interfaces have also been observed on simpler (non-polymeric) molecular systems, including hydrophilic and hydrophobic fluorescent dyes \cite{skaug2013intermittent}, as well as fluorescent ionic defects \cite{comtet2020direct,comtet2021anomalous,zhao2025defect}. This suggests the existence of potentially universal processes, independent of the macromolecular nature of the system, underlying such glassy energy landscapes at interfaces. Regardless of the exact underlying distribution, we can compute an average adsorption time, here of order $\langle \tau_\text{ads} \rangle \approx 100$ ms. Expressing this time as an activated barrier hopping scaling as $ \nu^{-1}\exp(-\Delta F/k_\text{B}T) $, with $\nu^{-1}\approx 10^{-13}$ s a molecular time taken as $h/k_\text{B}T$ with $h$ the Planck's constant, leads to an adsorption free energy $\Delta F$ of 0.81 eV or 32 $k_\text{B} T$, to be compared to the strength of a hydrogen bond, of the order of 2 $k_\text{B} T$

\paragraph{\textbf{Flow-induced dynamics.}}
\begin{figure}[htb!]
\centering
\includegraphics[width=170 mm]{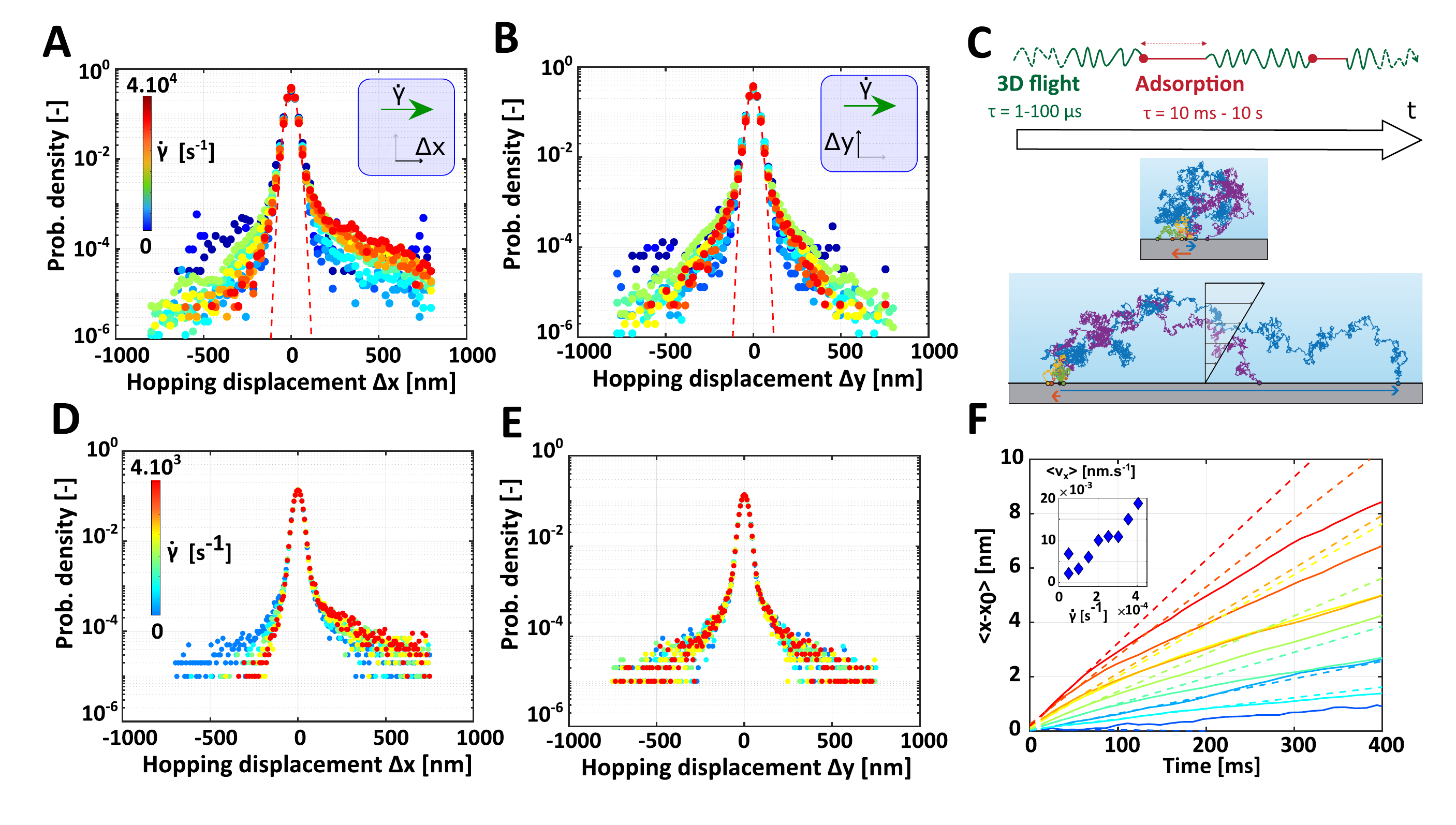}
\caption{\textbf{Shear-induced chain dynamics at hydrophilic interface - }\textbf{(A-B)} Displacement distributions (A) along $\Delta X$ and (B) transverse $\Delta Y$ to the flow direction at a hydrophilic surface for increasing shear rates, color-coded from blue ($0 \text{ s}^{-1}$) to red ($4 \times 10^4 \text{ s}^{-1}$). \textbf{(C)} Modeling of the experimental situation : diffusion in the near surface leads to adsorption during a random waiting time, before desorption and jump to another site. Representative jumping trajectories along and transverse to the flow direction are represented (see Materials and Methods). Corresponding hopping displacements $\Delta x$ and  $\Delta y$ are shown for two trajectories (blue and orange arrows).
\textbf{(D-E)} Simulation results for transverse (D) and longitudinal (E) motion, under  increasing shear rate, color-coded, from blue ($0 \text{ s}^{-1}$) to red ($4 \times 10^3 \text{ s}^{-1}$). \textbf{(F)} Averaged displacement $\langle (x(t_i+\Delta t)-x(t_i) \rangle_{t_i}$ as function of lag time $\Delta t$, shown for increasing shear rate (blue to red). Dotted lines represent linear fits on the first 40 ms. The evolution of the averaged velocity of the macromolecules at the interface extracted from the linear fit is shown in inset.}
\label{fig:Phile2}
\end{figure}

We now examine Figure~\ref{fig:Phile2} to analyze in detail how shear rate affects individual molecule displacement distributions, both transverse and longitudinal to the flow direction. For clarity, these distributions are presented for an effective lag time of $n \tau_0 = 120$\,ms. In Figure~\ref{fig:Phile2}A--B, we compare the transverse and longitudinal displacement distributions relative to the flow direction for shear rates ranging from 0 to $5 \times 10^4$\,s$^{-1}$. The gaussian-like central peak of both longitudinal and transverse distributions remains unchanged with increasing shear rate. This observation is consistent with the independence of the desorption time with the flow (Figure~\ref{fig:Phile1}C), which suggests that shear flow has no influence on macromolecules in the adsorbed state. In contrast, the longitudinal distributions show a marked dependence on the hydrodynamic flux (Figure~\ref{fig:Phile2}B). The long-distance tails appear indeed significantly altered by the flow: the probability of downstream displacements ($\Delta x > 0$) increases, whereas the probability of upstream displacements ($\Delta x < 0$) decreases. This suggest that the flow results in biased transport by influencing primarily the transport in the desorbed state: once desorbed, the chain is convected by the flow and tends to readsorb downstream.

To characterize in more details the effect of hydrodynamic flow on these interfacial processes, we turn to Brownian dynamics simulations. In order to allow a comparison with our experiments at the given sampling time, we consider explicitly, as schematically represented in Figure~\ref{fig:Phile2}D, the two important transport steps \cite{zhao2025defect}: \textit{(i)} 3D flights during which the polymer undergoes 3D Brownian motion in the bulk before readsorbing (Figure~\ref{fig:Phile2}D, green), leading to an apparent hopping step and \textit{(ii)} adsorption times (Figure~\ref{fig:Phile2}E, red), during which the molecule remains immobile at the interface.

We focus first on the dynamics in the freely desorbed state. We employ a simplified model, where the macromolecule is described as a point-like colloidal particle, neglecting internal conformational changes of the chain. This assumption is justified by the low value of the Weissenberg number $\text{Wi}=2\dot \gamma  \tau_\text{chain} \ll 1$, where $\tau_\text{chain}$ is the longest polymer relaxation time, evaluated from its Zimm motion (see Materials and Methods Section \myref{sec:MM_chemistry}). As the polymer undergoes 3D motion in the bulk, we described its dynamics by overdamped Langevin equations in the three spatial dimensions, with an additional term accounting for the convective flux along the $x$ direction (see Materials and Methods Section \myref{sec:MM_simu}). As schematically represented by the flow profile in Fig.~\ref{fig:Phile2}E, this convective effect is characterized by a Couette-like velocity profile, $v_x(z)=\dot{\gamma} \cdot (z - \lambda_\text{slip})$, where $\dot{\gamma}$ is the shear rate at the interface and $\lambda_\text{slip}$ the interfacial slip length, taken as $\lambda_\text{slip} \approx 0$ for this hydrophilic interface.
The equation of motion for the desorbed polymer writes accordingly:
\begin{equation}
\begin{split}
\dot x(t) &= \dot \gamma\cdot (z(t)-\lambda_\text{slip}) + \left( \frac{2k_BT}{\xi}   \right)^{1/2} w_x(t) \\
\dot y(t) &=  \left( \frac{2k_BT}{\xi}   \right)^{1/2} w_y(t) \\
\dot z(t) &=  \left( \frac{2k_BT}{\xi}   \right)^{1/2} w_z(t) 
\end{split}
\end{equation}
with $w_{x,y,z}$ the stochastic Langevin force accounting for random Brownian force, modeled as a white Gaussian noise with zero mean and delta-correlated variance.
 
Solving for individual trajectories, we show in Figure~\ref{fig:Phile2}F  three representative trajectories respectively in the $x$ and $y$ directions, and for a shear rate $\dot{\gamma} = 10^3$\,s$^{-1}$.  The Brownian particles representing our polymers start at an infinitely small distance to the interface, following which they explore the upper half space $z>0$ under a 3D exploration before eventually crossing back the interface, a condition which we treat as a re-adsorption (neglecting the presence of any barrier to re-adsorption). These elementary hopping distances  $\delta x$ and $\delta y$ are defined as the range traveled from the origin until this first re-encounters with the interface and characterized in Figure~\ref{fig:Phile2}C by the colored arrows. Along the transverse direction in $y$ the trajectories appear symmetrically distributed around the origin, while they show a clear flow-induced bias in the longitudinal $x$ direction. Note that the trajectory in $z$ being equivalent to a 1D random walk in the half space, the probability to reach back the interface is equal to 1 at very long time, i.e. all molecules eventually return to the interface.
%, and the time spent in the subsurface is power-law distributed, characteristic of a first passage time process, with typical time-scales of the order of {\color{red} XXX\,$\mu$s} for hopping steps of the order of few micrometers. Such short timescale are consistent with hopping steps appearing as instantaneous displacement at our 10 ms exposure time.
Upon re-adsorption, we assume our molecules to remain adsorbed to the interface for a randomly distributed waiting time, following an ad-hoc power-law distribution $P(\tau)\sim \tau^{-2}$ and a preset adsorption time $\tau_\text{m}$, as characterized in Fig. ~\ref{fig:Phile1}C. For comparison with our experimental distributions, these interfacial random walks are then directly sampled at the experimental sampling time.

The resulting distributions are then reported in Fig.~\ref{fig:Phile2}D-E, for increasing interfacial shear rates $\dot \gamma \in [0-10^3]$ s$^{-1}$ and mean adsorption time $\tau = 100$ ms, providing satisfactory qualitative agreement with our experiments. In particular, we note that our simulations capture the progressive asymmetry induced by the flow and show a stronger effect of the flow on the up-stream distribution ($x<0$), compared to the one down-stream ($x>0$). Anticipating our discussion below, the reduced effect of the flow on the downstream displacements can be understood from the already long-ranged power-law distributed displacements occurring already in the equilibrium case.

Several quantitative discrepancies between our simulations and experiments should however be highlighted. Considering first the transverse direction, while scaling of the long distance arms seem to be relatively well described by a power-law distribution as $P(|\Delta x|)\sim |\Delta x|^{-\alpha}$ in both experiments and simulations, the scaling power varies from $\alpha\approx 3$ in the experiments, to $\alpha \approx 2$  in the simulations  (see Supplementary Figure \ref{fig:simu_LongDistance}). Yet both these values are found in the same range as the previously reported scalings, ranging from $\alpha=2.5$ found experimentally \cite{yu_single-molecule_2013}, to $\alpha=2.3$ through lattice Monte Carlo simulations \cite{skaug2013intermittent}. Note that more advanced analytical solutions have also been used to describe experimental displacement distribution \cite{chechkin2012bulk,skaug_single-molecule_2014}.

Second, the effect of the flow in the longitudinal direction seems also to be stronger in our simulations than in our experiments, with a difference in the range of associated flow rates needed to reproduce the experimental behavior (with the maximal interfacial shear rate set to $10^3$ s$^{-1}$ in our simulations, compared to $10^4$ s$^{-1}
$ in the experiments. These various discrepancies are not unexpected, given the simplicity of our model. It would thus be interesting to include additional contributions, such as explicit reabsorption probability, distance-dependent friction with the wall \cite{alexandre2023non}, as well as the presence of equilibrium and non-equilibrium forces influencing near-surface dynamics \cite{sendner_shear-induced_2008, sing_non-monotonic_2011, ma_theory_2005}.

%\paragraph*{\textbf{Sub-convective transport}}
Returning to Fig.~\ref{fig:Phile2}F, we evaluate the ensemble transport by reporting the evolution of the displacement along the flow, $\langle x(t_i+\Delta t) - x(t_i) \rangle_{t_i}$, as a function of the delay time $\Delta t$. The hydrodynamic flow clearly increases the transport rate (from blue to red). Extracting an averaged velocity $v$ from linear fits of the displacement with time (Fig.~\ref{fig:Phile2}C, dotted lines) , we indeed observe a linear increase with the shear rate as $v\sim \alpha \dot \gamma$, where $\alpha \approx 1$~\text{pm}.
However, a striking observation is that these in-flow displacements exhibit a sub-linear scaling with time, $\langle x(t) - x(t=0) \rangle \sim t^a$, with $a \approx 0.8$. Such behaviors are a signature of sub-convective transport, analogous to the sub-diffusive growth of the mean-squared displacement observed in strongly adsorbing systems, which we attribute to the presence of largely distributed adsorbing periods. The observation of such a peculiar convective regime in strongly adsorbing systems is particularly interesting in the context of nanofluidic transport and filtration. It demonstrates that surface interactions can influence not only the transport rate but also its temporal scaling, influencing in profound way the transport kinetics.

\subsection{Dynamics on slippery hydrophobic surface.}

\paragraph{\textbf{Transverse displacement distribution of physisorbed chains.}}

\begin{figure}[htb]
\centering
\includegraphics[width=120 mm]{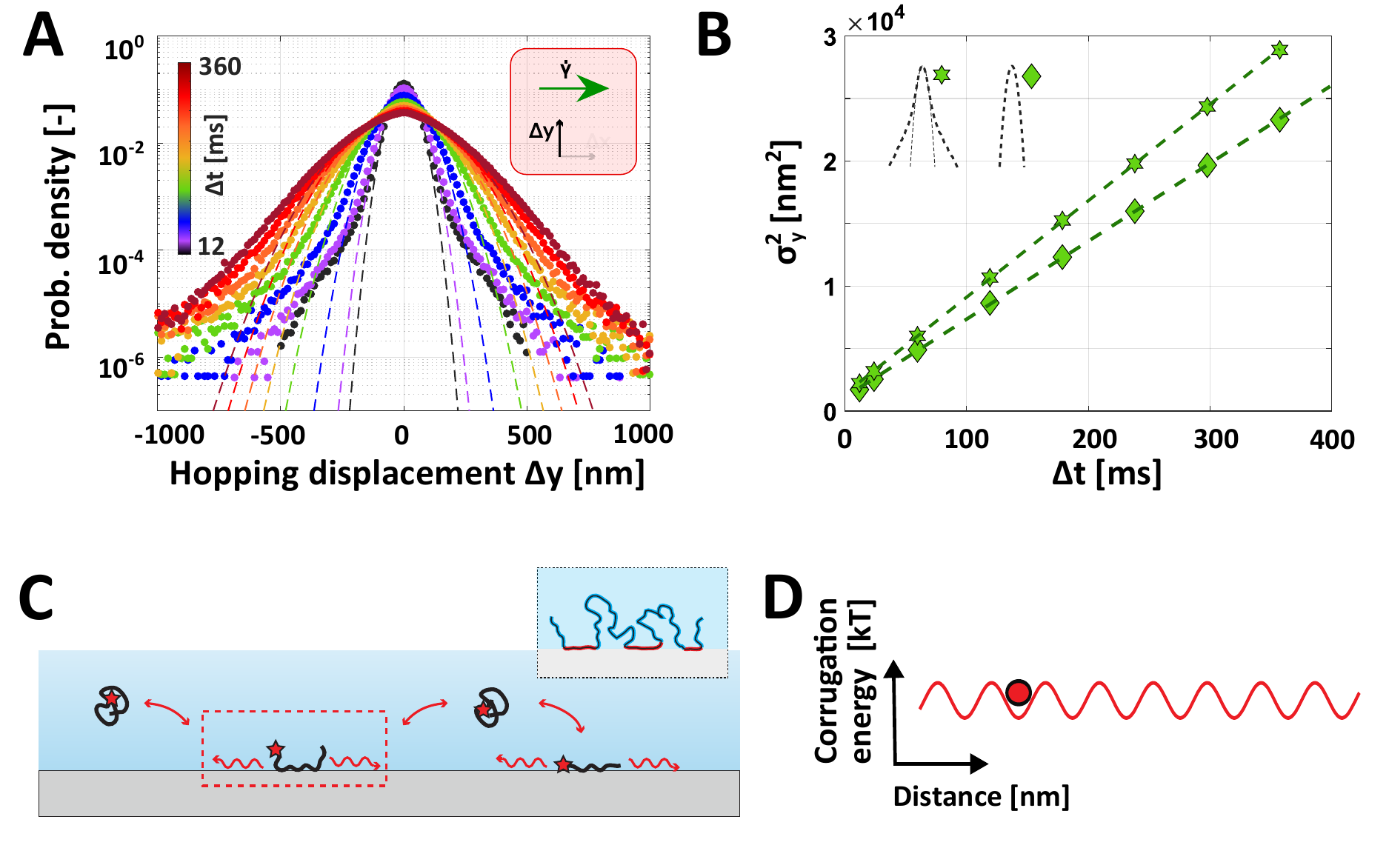}
\caption{\textbf{Polymer dynamics under flow at a slippery hydrophobic interface -} \textbf{(A)} Transverse displacement distributions along $y$ at an interfacial shear rate $\dot \gamma = 4\times 10^4$ s$^{-1}$, shown for increasing sampling times $\Delta t$. The dashed lines correspond to gaussian fits. \textbf{(B)} Evolution of the variance $\sigma^2$ with time, for the ensemble distribution $\sigma_\text{tot}^2$ and central gaussian part $\sigma_\text{gauss}^2$. The linear evolution with time characterizes a Fickian transport, with respectively $D_\text{tot} =38 \times 10^{-15}$ m$^2$.s$^{-1}$ and $D_\text{gauss} = 15 \times 10^{-15}$ m$^2$.s$^{-1}$. \textbf{(C)} Schematic of the interfacial transport mode, combining 2D interfacial diffusion with 3D hopping motion. As show in inset, transport in the physisorbed state is limited by the combination of friction with the solid wall (red) and with the liquid (blue). \textbf{(D)} Schematic representation of the smooth potential of hydrophobic surface allowing for the chain to diffuse laterally while remaining physisorbed. }
%{\color{red} Add a small schematic to highlight what $\sigma_\text{tot}$ and $\sigma_\text{gauss}$ are.}
\label{fig:Phobe1}
\end{figure}
\begin{figure}[htb]
\centering
\includegraphics[width=180 mm]{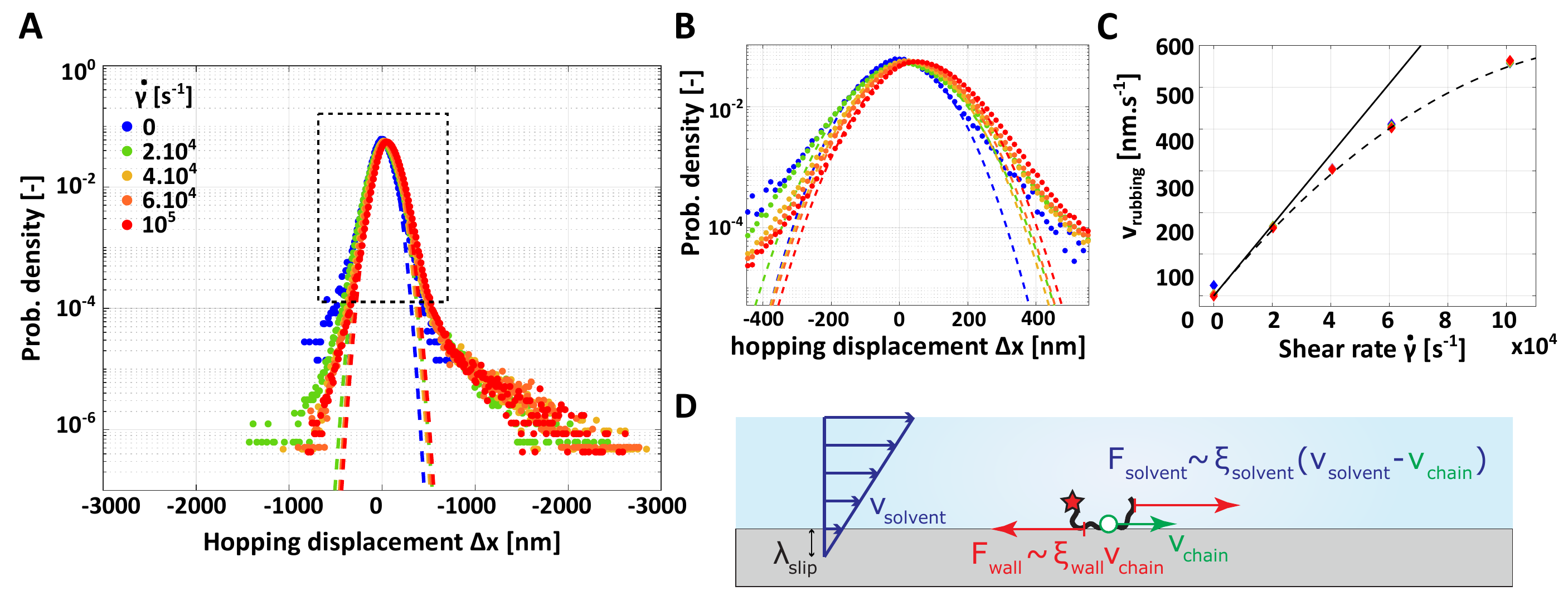}
    \caption{\textbf{Rubbing motion of physisorbed chains.} \textbf{(A)} Hopping displacement distribution along $ x$ for an hydrophobic surface under an interfacial shear rate between $0$ (blue) and $10^5$ $\text{s}^{-1}$ (red). The interval time is 100 ms. \textbf{(B)} Zoom on the central peak of the displacement distribution on hydrophobic surface along the flow (A). The dotted lines account for the Gaussian fit of the central peak. \textbf{(C)} Mean velocity of the chains along the flow in the physisorbed mode with respect to the shear rate. The values are calculated from the mean of the Gaussian fits shown in (B) divided by the considered time interval. The plain black line is a linear fit and the dotted line is a guide to the eyes.  \textbf{(D)} Schematic of the force balance of a single chain adsorbed at the interface (see text for details).% A slip boundary condition is considered as the surface is hydrophobic. A single tagged chain is physisorbed at the interface and has a velocity $\text{v}_{\text{chain}}$. Two main forces are applied on the chain: one through the solvent, i.e. the water molecules collisions on the chain, defined by $\textbf{F}_\textbf{solvent} \sim \xi_\text{s} (v_{\text{solvent}} - v_{\text{chain}})$ and the other by the wall, function of the interaction between the chain and the surface, given by $\textbf{F}_\textbf{wall} \sim  \xi_\text{w} v_{\text{chain}}$.
    }
\label{fig:Phobe2}
\end{figure}

We now turn back to the case of hydrophobic surfaces, for which the representative trajectories, reported in Fig~\ref{fig:Traj}D-F, highlighted already at a qualitative level a much more continuous transport motion that in the hydrophilic case (Figs.~\ref{fig:Traj}D-F). We focus in Figs.~\ref{fig:Phobe1} and~\ref{fig:Phobe2} on a quantitative analysis of the statistical properties of these distributions.
%a similar analysis framework, 
%While the distributions retain a clearly non-Gaussian character compared to the hydrophilic case, several key differences must be highlighted.
We first examine the displacement distributions transverse to the flow, which we report in Fig.~\ref{fig:Phobe1}A  for increasing sampling times (the data corresponds to a fixed shear rate $\dot \gamma = 4\times10^4$ s$^{-1}$, with similar evolutions in the equilibrium case, see Fig.~\ref{fig:DistribEq}).
Similar to the hydrophilic case (Fig.~\ref{fig:Phile1}), these displacement distributions present a clearly non-Gaussian character, with a combination of Gaussian-like distributions for short displacements, completed by decaying tails at longer distances. However, several major differences must be highlighted compared to the previous hydrophilic case.

In particular, the central Gaussian mode is larger than in the hydrophilic case at $\Delta t = 10$~ms, and shows a clear broadening with increasing sampling time, suggesting that macromolecules in the adsorbed state can maintain a significant lateral mobility. This broadening is quantified in Fig.~\ref{fig:Phobe1}B, where we report the evolution with the sampling time of the variance of this central Gaussian mode $\sigma_\text{gauss}^2$(green squares). The clear linear growth of $\sigma_Y^2$ with time (Fig.~\ref{fig:Phobe1}B, dotted line) demonstrates that the dynamics in the adsorbed state is associated to Fickian diffusion, while the observed Gaussian distributions show that the underlying regime is related to classical Brownian-like transport. These observations are thus consistent with a peculiar mode for macromolecular transport associated to a bidimensial Brownian-like diffusion in a weakly physisorbed state. The underlying diffusion coefficients, relating the Gaussian variance $\sigma_\text{gauss}^2$ to sampling time $\Delta t$ through  $\sigma_\text{gauss}^2 = 2 D_\text{2D}\Delta t$, lead to a bidimensional diffusion coefficient $D_\text{2D} = 15 \times 10^{-15}$~m$^2$.s$^{-1}$.

The recovery of these idealized Brownian  profiles are remarkable for such solid surfaces. We are indeed unaware of previous reports of such smooth Brownian-like interfacial diffusion for synthetic polymers on solid surfaces, except for the peculiar case of interfaces between liquid PDMS and water~\cite{wang2015scaling}. Notably, this behavior demonstrates the smoothness of the interfacial corrugation profile generated by the self-assembly of the octadecyltrichlorosilane monolayer on these solid surfaces (Fig. 5D).

In this context, this interfacial diffusion coefficient in the physisorbed state can be expressed as
\begin{equation}
D_\text{2D}=\frac{k_\text{B} T}{\xi_\text{w} + \xi_\text{s}}
\label{eq:diffusion}
\end{equation}
 with $\xi_\text{w}$ and $\xi_\text{s}$ characterizing respectively wall-induced and liquid-induced friction (Fig.~\ref{fig:Phobe1}C, inset).

In parallel to transport in this physisorbed state, we can also identify in the distributions of Fig.~\ref{fig:Phobe1}, the occurrence of long-tails in the displacements, deviating from the central gaussian mode, and which we interpret as stemming from fast hopping through 3D transport, in a similar fashion as that evidenced in Fig.~\ref{fig:Phile2} on hydrophilic surfaces. The variance of the ensemble distribution $\sigma_\text{tot}^2$ follows a similar linear growth, where the relatively slight differences between  $\sigma_\text{tot}^2$ and $\sigma_\text{gauss}^2$ demonstrates in this situation relatively minor effects of these hopping steps on interfacial transport. Fig.~\ref{fig:Phobe1}C represents schematically the transverse/equilibrium interfacial transport mode on these surfaces, as a combination of in-plane physisorbed motion with fast out-of-plane hopping.

%\section{Discussion}

\paragraph{\textbf{Rubbing transport}}

We now focus, in Fig.~\ref{fig:Phobe2}, on the interfacial transport along the direction of the hydrodynamic flow on these hydrophobic surfaces. The corresponding longitudinal displacement distributions along the $x$-axis are shown for an effective sampling time of $n\tau_0 = 120$~ms and increasing shear rates, ranging from $0$ to $10^5$~s$^{-1}$ (Fig.~\ref{fig:Phobe2}A, from blue to red). A first observation is that applying a shear rate at the interface induces an asymmetry in the tails of the hopping distributions, as previously observed for hydrophilic surfaces (see Fig.~\ref{fig:Phile2}). Yet, another remarkable feature specific to these hydrophobic surfaces also arises on these distributions. As represented in the zoom-in of Fig.~\ref{fig:Phobe2}B, the central gaussian distribution indeed tends to drift towards positive values upon increasing the shear rate. 
This gaussian offset $\mu$ grows linearly with sampling time, allowing us to extract an average convective drift velocity $v$ as $\mu=v\Delta t$. % as shown in \textbf{\color{red} SI.SXX}.
We report in Fig. \ref{fig:Phobe2}C the evolution of this interfacial rubbing velocity, which can be approximated as growing linearly with the applied shear rate, as ${v}_{\text{rubbing}} = {a } . \dot{\gamma}$ with $a\approx 8$ pm (straight line) followed by a sub-linear evolution at higher shear (dotted line).%As schematically illustrated in Fig.~\ref{fig:Phobe2}D, we interpret this effect as stemming from a rubbing-like motion of the adsorbed chains, which are continuously dragged by the hydrodynamic flow while retaining their interfacial mobility.

To account for the drifting behavior of the physisorbed chains, we consider a simple balance of force on a chain adsorbed at the interface, as schematized on Fig. \ref{fig:Phobe2}D. Two forces are accordingly working against each other, namely a hydrodynamic frictional drag force ${F}_\text{solvent}$, dragging the chain in the flow direction, resisted by the frictional force induced by the solid boundary $F_\text{wall}$. In the framework of linear response theory, we assume a linear relation between these frictional forces and macromolecular velocity with $F\sim v$, here perfectly justified by the Gaussianity of the short-ranged displacements evidenced in Fig.~\ref{fig:Phobe1}.

In this framework, the solvent-induced frictional forces on the chain is expressed as:
\begin{equation}
\text{F}_\text{solvent} \sim \xi_\text{s} (v_{\text{solvent}} - v_{\text{chain}})
\label{eq:F_solvent}
\end{equation}
where $\xi_{s}$ is the solvent induced friction coefficient [$\text{kg.s}^{-1}$] and $v_{\text{solvent}} - v_{\text{chain}}$ is the relative velocity between the chain and the surrounding solvent molecules.

Meanwhile, the wall-induced friction writes as:
\begin{equation}
\text{F}_\text{wall} \sim  \xi_\text{w} v_{\text{chain}}
\label{eq:F_wall}
\end{equation}
where $\xi_\text{w}$ represents the friction coefficient of the polymeric chain with the wall surface.

At steady state, these two forces balance as $\text{F}_\text{solvent} \sim \text{F}_\text{wall}$ and we thus obtained from Eq. \ref{eq:F_solvent} and \ref{eq:F_wall} : 
\begin{equation}
    {v}_{\text{chain}} \sim \frac{\xi_\text{s}}{\xi_\text{s} + \xi_ \text{w}} {v}_{\text{solvent}} 
    \label{eq:Velocities_relation}
\end{equation}
which can be also expressed as: 
\begin{equation}
    {v}_{\text{chain}} \sim \frac{\xi_\text{s}}{\xi_\text{s} + \xi_ \text{w}} \lambda \dot \gamma
    \label{eq:Shear_rate_relation}
\end{equation}
where the interfacial solvent velocity is approximated as a linear function of the shear rate $\dot \gamma$, through an interfacial slip length expected of the order of $\lambda \approx 10$ nm on these surfaces \cite{audry_amplification_2010}, potentially enhanced by the slightly extended conformation of the chain at the interface.

This simple approach allows us to recover the linear relation between drift velocity and shear rate evidenced experimentally in Fig.~\ref{fig:Phobe2}, where the proportionality coefficient is set by the relative chain/wall and chain/solvent friction. Combining these results in out-of-equilibrium conditions (Fig.~\ref{fig:Phobe2}, Eq.~\ref{eq:Shear_rate_relation}), with the equilibrium measurements of the diffusion coefficient (Fig.~\ref{fig:Phobe1}, Eq.~\ref{eq:diffusion}), we can disentangle the respective values of these solid and liquid friction coefficients, which are extracted as: 
\begin{equation}
    \xi_{\text{w}} \approx 1.3\times 10^{-7} \text{ kg.s}^{-1}\\
    \text{ and }\\
    \xi_{\text{s}} \approx 1\times 10^{-10} \text{ kg.s}^{-1} 
\end{equation}
A few comments are in order regarding these values. First, while the value obtained for liquid friction depends on the actual value of the interfacial slip length, liquid friction is still found to be very close to the value $\xi_\text{s}^\text{bulk}= 1.4\cdot 10^{-10} \text{ kg.s}^{-1}$ expected in bulk (taking an averaged bulk diffusion coefficient of $4\cdot 10^{-11}$ m$^{-2}$.s$^{-1}$ for 20 kg.mol$^{-1}$ PEG at 25 $^\circ$C \cite{waggoner_1995_diffusion,wang2015scaling}). This similarity suggests that the presence of the interface has little effect on the coupling of the adsorbed chains with the flow, consistent with chains keeping an extended conformation at the interface \cite{skaug_single-molecule_2014}. Conversely a three order of magnitude increase of friction is observed due to the interaction with the solid wall. This wall-induced friction,  can thus be expressed equivalently in terms of an equivalent "wall viscosity", reaching here values of the order of $1 \text{ Pa.s}$. Building upon models for adsorbate transport \cite{persson2013sliding} (Fig.~\ref{fig:Phobe1}D), we can express the friction coefficient as $\xi \sim k_B T/(\nu \lambda^2)  \exp(\Delta F/k_\text{B}T)$, with $\nu = k_BT/h \approx 10^{-13}$ s the previously defined molecular attempt frequency, and $\lambda\approx 0.3$ nm a typical corrugation length of molecular size, leading to $\Delta F\approx 0.24$ eV, significantly smaller than the mean corrugation energy found on the hydrophilic surface, see Fig.~\ref{fig:Phile2}. Finally, the sub-linear response at high shear rate can be interpreted as stemming from non-linear effects in the friction coefficient  

\subsection{Down to single-chain friction}

\begin{figure}[htb]
\centering
\includegraphics[width=180 mm]{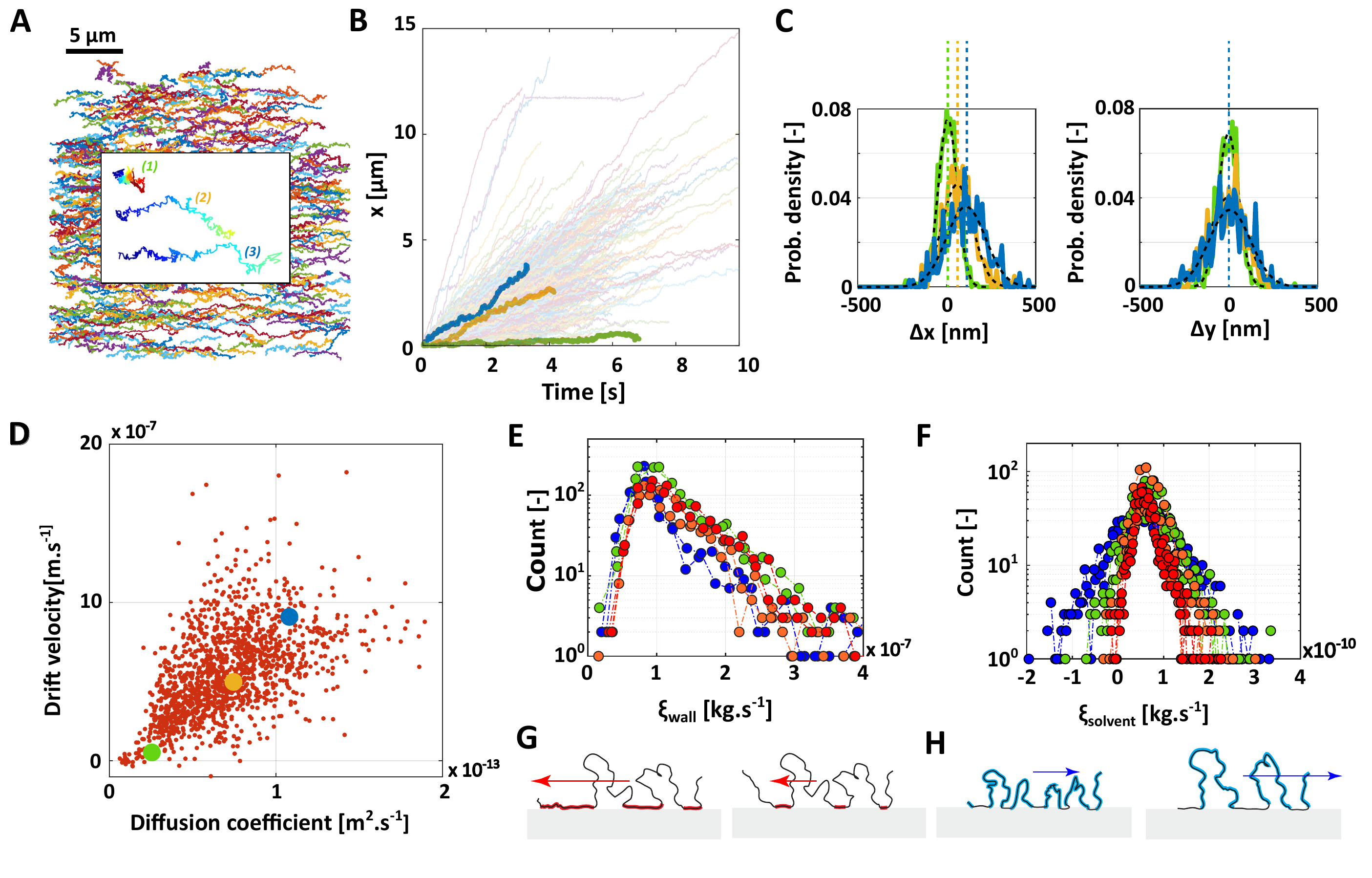}
\caption{\textbf{ Analysis of individual chain rubbing at an hydrophobic interface -} \textbf{(A)} Plot of individual trajectory map at $\dot \gamma = 10^5 \text{ s}^{-1}$. Only trajectories larger than 300 frames are shown. The inset highlights three representative trajectories, which exhibit distinct drift dynamics. The color-map in the trajectories correspond here to the same time-base. \textbf{(B)} Cumulative displacement along the flow for the trajectories shown in (A). \textbf{(C)} Displacement distributions of the three trajectories. Dotted lines display the  Gaussian fit of the distributions, and vertical line their mean. \textbf{(D)} Scatter plot showing correlation between rubbing speed and diffusion coefficient at the scale of the chain population. \textbf{(E)} Distribution of  $\xi_{\text{wall}}$ for increasing shear rate conditions from $0 \text{ s}^{-1}$ (in purple) to 10 $\times 10^4 \text{ s}^{-1}$ (in red). \textbf{(F)} Distribution of  $\xi_{\text{solvent}}$ for increasing shear rates. \textbf{(G - H)} Schematic conformation of adsorbed chain undergoing low (left) or high (right) friction respectively with the wall (G) or with the solvent (H).}
\label{fig:Individual_Rubb}
\end{figure}

Our approach has successfully characterized the peculiar frictional interactions between adsorbed polymeric chains and solid surfaces. However, this approach remained limited to averaged behaviors at the scale of the chain ensemble. In the following, we leverage our single-molecule approach to focus on the behavior of individual chains within the population.

We first focus on the dynamics at the hydrophobic surface at the highest shear rate of \(10^5 \, \text{s}^{-1}\).
We represent in Fig. \ref{fig:Individual_Rubb} A a spatial maps of individual trajectories (focusing on those longer than 300 frames for clarity). 
The inset highlights three representative trajectories, each exhibiting distinct drift mobility behaviors. To characterize the heterogeneity of these behaviors at the population scale, we report in Fig.~\ref{fig:Individual_Rubb}B the displacements along the \(x\)-axis as a function of time for all trajectories.
Two key features emerge from this analysis:
first, the drift velocities of individual chains remain approximately constant over the entire trajectories, as evidenced by the linear scaling of \(x(t)\) with time;
second, these individual velocities are broadly distributed across the chain population, as characterized by the spread of these trajectories.
The displacements for the three representative trajectories highlighted in Fig.~\ref{fig:Individual_Rubb}A are shown in green, yellow, and blue, respectively.

To further analyze these behaviors at the single chain level, we present in Fig.~\ref{fig:Individual_Rubb}C the hopping displacements \(\Delta x\) and \(\Delta y\) for these three representative trajectories, along the longitudinal (\(x\)) and transverse (\(y\)) directions, respectively.
Each distribution is accompanied by its associated Gaussian fit (black dotted line).
As expected, the fastest chain (blue distribution) exhibits the largest offset (indicated by the vertical blue dotted line in the \(\Delta x\) distribution), followed by the yellow and green distributions.
However, a second notable feature is also apparent in these plots: fast drifting chains (blue) appear to exhibit a significantly broader width in their Gaussian-like displacement distribution. This effect can be seen in both the transverse and longitudinal directions, and suggest an enhanced diffusive mobility for fast drifting chains. %Respective values for drift velocities and diffusion coefficients for the three trajectories amount respectively to {\color{red}1.4, 4 and}.

Having observed this correlation between drift and diffusive behavior in the three representative cases, we examine how this relationship extends to the entire population. In order to quantitatively evaluate these joint equilibrium and out-of-equilibrium properties, we fit the distribution of hops for individual trajectories with a non-centered gaussian, of mean $\mu$ and width $\sigma$. Individual drift velocities and diffusion coefficients can be simply calculated respectively as \(v_\text{chain} = \mu/\Delta t\) and $D_\text{ind} = {\sigma^2}/{2 \Delta t}$, where \(\mu\) and \(\sigma\) are the transverse offset and mean of the gaussian fit of the displacement distribution and \(\Delta t\) is the sampling time.
%Additionally, for each trajectory, we extract an individual chain diffusion coefficient, using the width  of the Gaussian fit of the displacement distribution, as 

As shown in Fig.~\ref{fig:Individual_Rubb}D, the correlation reveals a robust and consistent relationship between these two quantities for the entire chain population, revealing a robust and consistent relationship.
This correlation between equilibrium and out-of-equilibrium transport coefficient is remarkable and can be understood through a simple argument: within our frictional framework presented in Fig.~\ref{fig:Phobe2}, both parameters are strongly influenced by the wall-induced friction coefficient, \(\xi_\text{w}\).
Specifically, in the limit \(\xi_\text{w} \gg \xi_\text{s}\), we find \(D_\text{ind} \approx k_\text{B}T/\xi_\text{w}\) and \(v_\text{ind} \approx \lambda \dot{\gamma} \xi_\text{s}/\xi_\text{w}\).
Thus, the observed correlation is a signature of the dominant role of wall friction in governing the underlying interfacial dynamics, while the remaining spread is attributed to variations in the solvent friction coefficient $\xi_\text{s}$.

%Rewriting the simple for balance expressed previously, this point is consistent with an approximately constant liquid friction coefficient $\xi_\text{s}$, leading to 
 %\begin{equation}
  %  v_\text{chain} \sim  \xi_\text{s} \frac{\dot\gamma \lambda}{k_\text{B} T} D
%\end{equation}

Based on this individual analysis, we can now extract the friction coefficients associated with individual trajectories and plot their distributions in Fig.~\ref{fig:Individual_Rubb}E--F. As expected, these two coefficients, \(\xi_{\text{s}}\) and \(\xi_{\text{w}}\), are centered around average values of \(1 \times 10^{-10} \, \text{kg} \cdot \text{s}^{-1}\) for \(\xi_{\text{s}}\) and \(1 \times 10^{-7} \, \text{kg} \cdot \text{s}^{-1}\) for \(\xi_{\text{w}}\), consistent with those obtained from the ensemble-average approach in Fig.~\ref{fig:Phobe2}. Yet the distributions show also a significant spread around these averaged values, reflecting the heterogeneity in the frictional interactions of the chains.

As represented in Fig.~\ref{fig:Individual_Rubb}G, H, this spread can be associated to distinct conformations in the adsorbed state, assuming the chains to adopt a loop-and-train configuration at the interface \cite{manciu_loops_2011}. For the solid friction coefficient, a tentative picture could account for a linear scaling with the number of adsorbed trains, characterized by the red region in Fig.~\ref{fig:Individual_Rubb}G, with $\xi_{\text{wall}} \sim N_{\text{train}} \xi_{\text{monomer}}$, where $\xi_{\text{monomer}}$ represents the monomeric friction coefficient with the surface. Likewise, the polymer/solvent friction coefficient should depend upon the three-dimensional conformation of the polymer chain near the surface, specifically through the number of adsorbed loops, as well as their spatial extension from the interface (Fig.~\ref{fig:Individual_Rubb}H). In particular, extended loops within the interface would be expected to result in comparatively higher friction coefficient \(\xi_{\text{solvent}}\).

With this picture in mind, it is remarkable that the value of these frictional coefficients seems to be well conserved over the entire observations time-scale of up to tens of seconds as characterized by the constant linear drift shown in Fig.~\ref{fig:Individual_Rubb}B. 
This behavior could be attributed to intrinsic polydispersity in chain length, with variations in the friction coefficient rather set by dispersity in chain length. Yet, such an explanation seems inconsistent with the low polydispersity of our sample (See Materials and Methods \ref{MM_dispersity}). These heterogeneity in the frictional behavior thus rather suggest that interfacial conformation is conserved once the chains adsorb to the interface. 
Taking the characteristic end-to-end distance in the adsorbed configuration as $R_0\approx 13$ nm (evaluated here from the one in bulk) and the solid friction coefficient as $\xi_\text{w} \approx 1.3\cdot 10^{-7}$ kg.s$^{-1}$ amounts to a characteristic time for interfacial conformation change equal to $R_0^2 \xi_\text{w}/k_\text{B}T \approx 5$ ms, inconsistent with the large frictional memory effect evidenced here. Further understanding of the mechanisms underlying this extended interfacial conformational stability opens exciting avenues for further investigation.

\subsection{Conclusion}
By coupling wide-field single molecule localization microscopy, transport on model surfaces in microfluidic settings and detailed statistical analysis, we investigated at the single macromolecular level, the role of surfaces and the effect of hydrodynamic flows on the dynamics of solvated polymeric chains adsorbed at a solid/liquid interface. Tracking chain trajectories at high spatial and temporal resolution allowed us to investigate the statistical properties of these interfacial random walks, and reveal a marked dependence of interfacial transport on surface properties. On hydrophilic glass surfaces, we recovered at equilibrium an heterogeneous and strongly non-Brownian surface diffusion mode for individual chains, which alternate between adsorbed states and long desorption-mediated jumps through the solvent. The distribution of adsorption time further follows peculiar power-law scaling reminiscent of glassy energy landscape. Oppositely, hydrophobic surfaces obtained by self-assembly of monolayers of long alkyl chains are characterized by a much smoother interfacial potential, with interfacial motion characterized by classical Brownian motion associated with bidimensional diffusion of physisorbed interfacial chains, occasionally combined with transient and discontinuous desorption-mediated jumps.

The symmetry-breaking effect of the flow leads to a skewed distribution of interfacial displacements, showing again strong dependence on the nature of the surfaces. On sticky hydrophilic surfaces, the hydrodynamic flow does not affect the chain motion except for an advective effect during solvent-mediated flights. Yet, during this transport, the presence of strong adsorption steps lead to a peculiar sub-convective transport dynamics in the flow direction. Our experimental displacement distribution are here well-captured by simple simulations of the biased flow-induced motion of Brownian particles close to a solid surface.
On slippery hydrophobic surfaces, we  evidence instead a peculiar regime of mixed solid and liquid macromolecular friction, whereby the adsorbed chain rubs on the solid wall while being continuously dragged by the near-surface hydrodynamic flow. In this case, joint analysis of equilibrium and out-of-equilibrium transport thus allows to finely disentangle chain/wall and chain/solvent molecular frictional interactions. Finally, we analyze the trajectories at the level of individual chains within the population. Behind the population averaged behavior, these measurements unveil a broad distribution of friction coefficients, which we attribute to a diversity of interfacial conformation with sluggish reconfiguration timescale. These direct observations of the out-of-equilibrium flow-induced transport of interfacially adsorbed polymers, bring a new molecular vision of macromolecular friction and adsorbate/surface interaction at flowing solid/liquid interfaces, and provide a foundation for the development of molecular models of the macroscopic behavior of interfacial soft matter \cite{chaudhury1999rate,blake2006physics,singh2011steady}.

Our observations and findings open up several exciting avenues and perspectives for the observation of soft matter dynamics and interfacial friction at the molecular scale. In the context of polymer physics and physical chemistry, it would be exciting to further investigate the role of molecular weight, chain architecture as well as solvent quality and surface affinity. These parameters would allow to probe in more details the role of chain conformation  on friction of the macromolecules respectively with the solid surface and with the solvent. Our approaches could further be extended towards denser semi-dilute or even melt regimes. In these conditions,  hydrodynamic or elastic correlations between chains could take place, and a stronger coupling between single-molecule behavior and the macroscopic ensemble response should arise. In the context of mechanochemistry and interfacial tribology \cite{gnecco2007fundamentals}, our approach has the potential to reveal molecular-scale force-induced response of soft materials, beyond the usually assumed ensemble-average Arhenian-like activation kinetics. Finally, beyond polymer physics, the ability for direct visualization of interfacial dynamics in out-of equilibrium systems open broad perspectives for molecular transport at interfaces \cite{alexandre2022stickiness} and the direct exploration of soft matter, liquids at interfaces and in confinement \cite{schlaich2025theory} and nanofluidics \cite{kavokine2021fluids} at the molecular-scale.

\section{Acknowledgments}

J.C. acknowledge funding from the ANR (grant ‘GUACAmole’ ANR- 22-CE06-0003-01), and Sorbonne University for funding the PhD of M.V. through a "Scientific Doctoral Policy" contract. This project has also received financial support from the Ile-de-France Region in the framework of ‘DIM Respore’ and ‘DIM MaTerRE’, the CNRS through the MITI interdisciplinary programs and from the Carnot Institute ’IPGG Microfluidique’. This work benefited from the technical contribution of the joint service unit CNRS UAR 3750 and was supported by UAR 2063 Workshop for Advanced Research Support.

J.C. acknowledges the encouragement and support of Costantino Creton. We thank  Lionel Bureau, Joshua McGraw, Thomas Salez for interesting discussions and feedback. We thank Mohamed Hanafi for help with the GPC measurement, Léa Gaonac'h, Josh McGraw and Emilie Verneuil for advice with the silanization procedure, as well as Nadège Pantoustier, Atul Sharma and Alex Cartier for advice on the grafting procedure.

\section{Supplementary Materials}

\begin{enumerate}
    \item \textbf{Supplementary Video S1} - Movie of PEG dynamics on hydrophilic surface, shear rate $\dot \gamma=10^4$ s$^{-1}$ \label{itm:SI_hydrophile_shear}
    \item \textbf{Supplementary Video S2} - Movie of PEG dynamics on hydrophobic surface, shear rate $\dot \gamma=0$ s$^{-1}$\label{itm:SI_hydrophobe_noshear}
    \item \textbf{Supplementary Video S3} - Movie of PEG dynamics on hydrophobic surface, shear rate $\dot \gamma=10^4$ s$^{-1}$\label{itm:SI_hydrophobe_shear} 
    \item \textbf{Supplementary Informations} - PDF\label{itm:SI_PDF}
   
\end{enumerate}

\section{Materials and Methods}

\subsection{ Polymer chemistry and dynamics}\label{sec:MM_chemistry}

\subsubsection{Polymer chemistry and tagging}
We used solutions of poly(ethylene glycol) (PEG) chains as a model synthetic polymer system. Poly(ethylene glycol)--NH\textsubscript{2} was purchased from Biopharma PEG Scientific Inc. and used for fluorescent grafting without further purification. The polydispersity index of the sample was determined to be 1.070 via Gel Permeation Chromatography, with weight averaged molar mass of 23 000 g.mol$^{-1}$.%, with $M_n = 21494$ and $M_w = 22999$.

Ester-functionalized fluorophores (ATTO 488 and ATTO 643) were obtained from ATTO-TEC. They are bright, stable and hydrophilic fluorophores classically used for single-molecule imaging. Quantitative grafting of the fluorophores onto the polymer chains was achieved through an amidation reaction in a buffered aqueous solution (pH 8) with an excess of dye, following the manufacturer's instructions. The one-pot reaction was incubated in the dark at room temperature for 1 hour. The mixture was then purified using a SEPHADEX G25 desalting column (Sigma-Aldrich). Thin-layer chromatography confirmed the absence of free dyes in the functionalized PEG solution. The final concentration in the reaction batch was estimated using UV--visible spectroscopy. The adsorption of ATTO 488 (-1e charge) on hydrophilic and silanized glass surface was previously found to be negligible \cite{yu_single-molecule_2013}. ATTO 643 bears similar -1e negative charge and present high hydrophilicity and excellent water solubility, ensuring in a similar way the absence of unspecific surface binding.

The functionalized PEG solution was stored at 4\,$^\circ$C until use. Prior to experiments, the grafted chains were diluted at concentrations ranging from \( 10^{-10} \) to \( 10^{-13} \, \text{mol}\cdot\text{L}^{-1} \) in Trizma buffer (Sigma-Aldrich), with a pH set to 8.2. Tagged PEG concentration was around $10^{-12}$ mol.L$^{-1}$ and $10^{-10}$ mol.L$^{-1}$ respectively for the experiments on hydrophilic and hydrophobic surfaces.

\subsubsection{Polymer dynamics}

PEG chains considered here have a degree of polymerization of $\approx$ 523, and diffusion coefficient in bulk taken as $D_\text{bulk}\approx 4 \cdot 10^{-11} \text{ m}^2.\text{s}^{-1}$ \cite{wang2015scaling,waggoner1995diffusion}. Chains can be modeled as random coils, characterized by a Kuhn molar mass \( M_0 = 137 \, \text{g}\cdot\text{mol}^{-1} \), a Kuhn length \( b = 1.1 \, \text{nm} \), resulting in a number of Kuhn segments \( N_k = 146 \) \cite{rubinstein_polymer_2003}. Water acts as  a good solvent for PEG, yet short molar mass show $\theta$-solvent like scaling \cite{sukhishvili2002surface,wang2015scaling}, with Flory exponent $\nu=0.5$. The end-to-end distance in bulk is thus evaluated as \( R_0 = 13.3 \, \text{nm} \), and the Zimm relaxation time\cite{zimm_dynamics_1956} found to be $\tau_Z \approx 10^{-7}$ s. The Weissenberg $Wi = 2 \dot \gamma \tau_Z$ remains small $<2\cdot 10^{-2}$, even at the largest  shear rate $\dot \gamma\approx 10^5$ s$^{-1}$, ensuring the absence of shear-induced deformation of the chain in the bulk.

\subsection{Surface preparation, microfluidic chip fabrication and optical experimental set-up}

\subsubsection{Surface Treatment} \label{sec:MM_surface}
Borosilicate glass coverslips (\#~1.5) were purchased from VWR. They were cleaned with Piranha solution. The coverslips were then sonicated twice in ultrapure water and dried under an argon flow. They were reactivated by a 15 minutes UV-Ozone treatment right before use.

Hydrophobic surfaces were obtained by silanization with octadecyltrichlorosilane (OTS, ABCR, 99.7\% purity) via liquid-phase deposition in toluene. After piranha cleaning and rinsing, glass coverslips were re-activated under UV-Ozone, and immersed in a 0.25\% volume silane solution during 20 min. Coverslips were thoroughly rinced in toluene and water in a ultrasonic bath, dried under an argon flow and stored carefully.

Coated layer thickness was characterized through ellipsometry on a reference silicon layer, leading to a thickness of 2.5 nm, consistent with previously obtained values for OTS Self Assembled Monolayers \cite{wang2003growth}. AFM imaging evidenced a relatively homogeneous surface state, with RMS roughness of 400 pm, combined with occasional small islands of nanometric height and tens of nanometer width, attributable to sparse polycondensation.

Silanized glass surfaces showed some residual localized fluorescence under laser excitation. To prevent interference with our single fluorescent molecule signal, the imaging area was bleached for $\approx 10$ minutes at high power. When employing the red laser, this protocol led to irreversible bleaching, ensuring proper subsequent characterization of interfacial PEG dynamics. Hydrophobic surfaces were thus studied using ATTO 643 tagged PEG chains, while the dynamics of ATTO 488 tagged chains where probed on hydrophilic surfaces.

\subsubsection{Microfluidic chip Fabrication.}
\begin{figure}[htb!]
    \centering
    \includegraphics[width=0.5\linewidth]{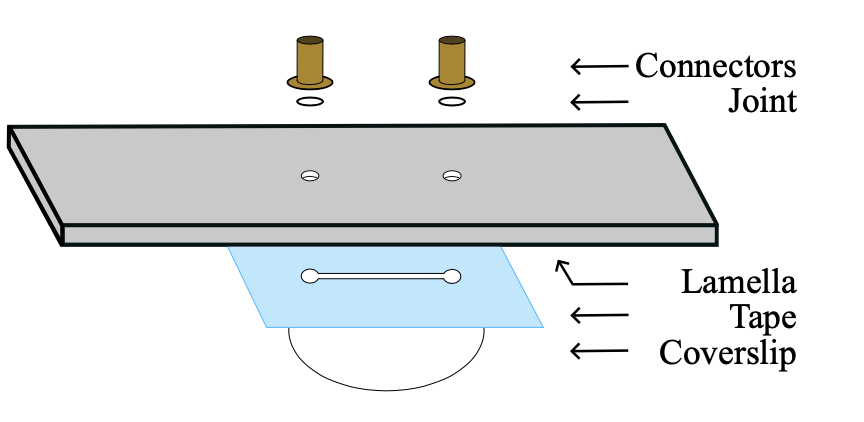}
    \caption{Microfluidic chip made with double sided tape. 
}
    \label{fig:SI_chip}
\end{figure}
Microscope lamellae of size 76 x 25 x 1 mm were purchased from VWR (ref. 631-1550). The inlet and outlet of the microfluidic chip were drilled using a CO\textsubscript{2} laser (C180II Desktop Laser Engraver). The lamellae were sonicated for 20~minutes in ultrapure water prior to use.

The microfluidic chip were built using research grade double side tape as 81 $\mu$m spacer (Adhesives Research , 90445). A straight channel shape was laser-cut in the tape (L = 20 mm, w = 0.5 mm), with 2 circles of diameter 1 mm at the edges. The chip was closed by the coverslip and the microscope lamellae on the other side. NOA Adhesive cured with UV allowed to fix microfluidic connectors from Idex H\&S (NanoPort for 1/16" OD tubing, Idex H\&S and Darwin Microfluidics)

The flow in the channel was regulated by an Elveflow pressure controller (OB1 controller, 0 to 2 bar). The PTFE microfluidic tubings were purchased from Darwin Microfluidics. Two Falcon tubes (50 mL, VWR) were used as the inlet and outlet reservoirs. The tubes and the reservoirs were changed in between each experiment to avoid any contamination by the previously used solutions.

\subsubsection{Description of the optical set-up.} \label{sec:MM_setup}We used a home made optical single molecule TIRF setup mounted on a vibration-damping optical table and build around an IX83 Olympus microscope body allowing for motorized control of the objective. The blue and red lasers used here are respectively a 488 nm laser diode from Oxxius (488 nm, 150 mW, REF L1C-561L-150-CSB-MPA-0-PP) and a 637 nm laser diode from Coherent (160 mW, OBIS 637 nm LX 160 mW). The laser beams are first combined and expanded before entering the TIRF setup. The 488 nm laser is first slightly expanded and cleaned with a  bandpass filter (ZET488/10x Chroma) centered around 480 nm. The 637 nm laser is simply cleaned with a pass band filter centered around 640 nm (Thorlabs FBH640-10). An achromatic quarter-wave plate (Thorlabs, AQWP05M ) transforms the linear polarization of the incident wave into a circular polarization and an achromatic beam expander with a fixed x10 magnification (Thorlabs, GBE10-A) broadens the beam.

The laser bems are finally focused on the back focal plane of an oil-immersion objective (x 100, ref. UPLAPO100XOHR) mounted on the Olympus IX83 microscope body. The objective has a high numerical aperture of NA = 1.5 to maximize photons collection. A manual stage allows to shift the beam with respect to the optical axis, allowing to reach illumination in TIRF mode.

Fluorescence intensity coming from the interface is collected by the objective, separated from the emission light through a quadband dichroic filter (Semrock Di03-R405/488/561/635-t1-25x36) and cleaned by an emission filter (Semrock FF01-446/523/600/677-25). The beam is then magnified by a factor 1.4 with an afocal system (two achromats doublet lenses from Qioptiq, $f_1 = 100 \text{ nm}$ and $f_2 = 140 \text{ nm}$) and projected onto the sensor of a cooled Electron-Multiplying Charged-Coupled Device camera (EMCCD, Andor iXon Ultra DU-897U-CS0, cooled down at -70°C). The associated projected pixel size takes a value of 114.3 nm ensuring proper sub-pixel localization while maintaining sufficient signal-to-noise. The Micro Manager software is used to control the lasers, camera and microscope body. Videos are recorded with an exposure time of 10 ms and frame interval of 12 ms. Typical acquisition length for one movie was 10'000 frames. The image size is $256\cdot256$ pixel$^2$, corresponding to a real field of view of $\approx 29$ $\mu$m wide. Illumination power at the back-focal plane ranges from $\approx$ 1.5 to 4 mW, leading to power densities $\approx 2-5 \cdot 10^{-4}$ mW.$\mu$m$^{2}$ (dividing the excitation power by $\pi\sigma_\text{OBJ}^2$ where $\sigma_\text{OBJ}$ is the gaussian width of the excitation profile).

%{\color{red} Give values for back focal plane power densities}.

%{\color{red}@MALO Ce serait pas mal d'ajouter le schéma optique du set-up que tu as fait pour ton manuscript, en enlevant le laser vert}

\begin{figure}[htb]
\centering
\includegraphics[width=150 mm]{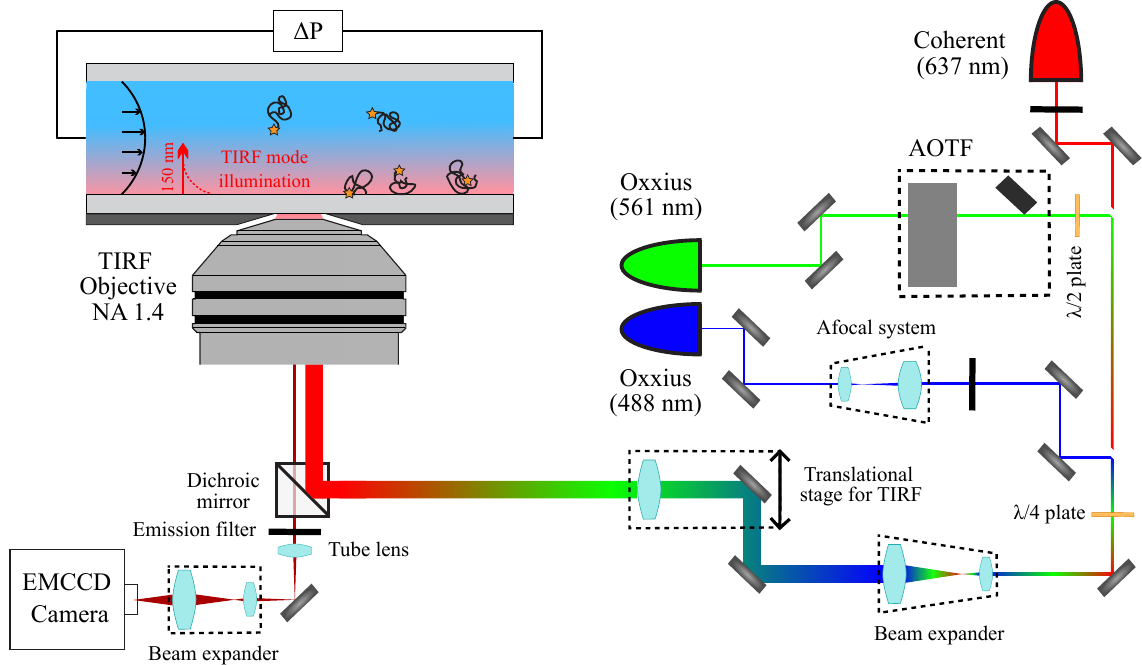}
\caption{\textbf{Optical pathway for our Single-Molecule Localization Microscopy (SMLM) set-up }}
\label{fig:SI_setup}
\end{figure}

\subsection{Data analysis}
\label{sec:MM_data}
\subsubsection{Localization}

Emitters are localized using the imageJ plugin, Thunderstorm \cite{ovesny2014thunderstorm}. Briefly, a wavelet filter is applied to each frame. Peaks are then fitted by 2D integrated Gaussian. Only emitters whose intensity peak at values at least 1.2 the standard deviation of the wavelet scale parameter are considered. This localization procedure allows to obtain localization tables for the spatial positions of active emitters at the surface of the flake at each time frames. Localization precision $\sigma_\text{loc}$ is calculated based on the Thompson-Larson-Webb Formula \cite{thompson2002precise}, as detailed by Ovesn\'y et al \cite{ovesny2014thunderstorm}. The leading order in localization precision scales as $\sigma_\text{loc}\sim\sigma_\text{PSF}/\sqrt{N_\phi}$, where $\sigma_\text{PSF} \approx 150$ nm is the standard deviation of the Gaussian fit of emitter's intensity (corresponding to a diffraction-limited spot fixed by the Point Spread Function with a full width at half-maximum of $\sim350$ nm), and $N_\phi$ is the number of photons emitted by the defect during the acquisition of one frame. Depending on experimental conditions, signal to noise ratio and brightness of emitters, the localization precision varies between 10 and 50 nm, with a median centered around 20 nm. 

Super-resolved images in Fig.~\ref{fig:Traj} are rendered as sums of all detected localizations, through a visualization algorithm based on averaged shifted histograms method with an effective pixel size of 11.5 nm.

\subsubsection{Tracking}
 In order to characterize the trajectories of fluorescently tagged chains at the surface of the flake, we applied a tracking algorithm on the emitter localization table detailed above. Briefly, active sites are identified as belonging to the same trajectory when they are present in consecutive frames, and are within a defined distance $l_\text{track}$. Given two successive frames, this assignment is obtained by minimizing the total squared distance, with non-associated particles penalized by the maximal squared displacement $l_\text{track}^2$. We used a version of the tracking algorithm developed by Blair and Dufresne and available online \cite{blair_dufresne}. $l_\text{track}$ was set respectively to 500 nm and 800 nm for the hydrophobic and hydrophilic surfaces, as a comprise between computational cost and false assignments. For biased flow experiments, a tilt correction was used to precisely identify the flow axis directions.

\subsubsection{Trajectory segmentation}

To discriminate between adsorbed and hopping steps, we define a cut-off ${l}_{\text{jump}}$ on the length of the displacement step $\Delta l = \sqrt{\Delta x^2 + \Delta y^2}$. All hopping steps longer than ${l}_{\text{jump}}$ are assimilated to hops and define a new adsorption period. We verified that the exact choice of ${l}_{\text{jump}}$ has a weak effect on the observed jumping displacement distributions. A distance ${l}_{\text{jump}}= 100$ nm was chosen for the hydrophilic surface, leading to the distribution shown in Fig.~\ref{fig:Phile1}C.

\subsection{Dispersity of the friction coefficient}\label{MM_dispersity}

To uncover whether the distribution of friction coefficient evidenced in Fig.~\ref{fig:Individual_Rubb} stems from intrinsic sample polydispersity or from conformational effects, we evaluate the polydispersity index (PDI) which would be expected based on the distribution of friction coefficient $P(\xi)$ measured in Fig.~\ref{fig:Individual_Rubb}. To do so, we assume a general power-law relation between chain friction with the solid $\xi$ and chain length $M/M_0$ as $\xi = \xi_0 (M/M_0)^\alpha$ where $M_0$ is the molar mass of the Kuhn monomer, $\alpha$ is a scaling factor and $\xi_0$ a monomeric friction coefficient.

The expected number and weight-averaged molecular weight can be expressed as follows, with $n_i$ the number of molecules with molecular weight $M_i$ and friction coefficient $\xi_i$.
\begin{eqnarray*}
M_\text{n}  = \frac{\sum n_i M_i}{\sum n_i} =\frac{\sum n_i M_0 (\xi_i/\xi_0)^{1/\alpha}}{\sum n_i} = \frac{M_0 }{\xi_0^{1/\alpha}}\frac{\sum n_i \xi_i^{1/\alpha}}{\sum n_i}% = \frac{M_0 }{\xi_0^{1/\alpha}} \xi_n
\end{eqnarray*}

and
\begin{eqnarray*}
M_\text{w}  = \frac{\sum n_i M_i^2}{\sum n_i M_i } =\frac{\sum n_i M_0^2 (\xi_i/\xi_0)^{2/\alpha}}{\sum n_i M_0 (\xi_i/\xi_0)^{1/\alpha}} = \frac{M_0 }{\xi_0^{1/\alpha}} \frac{\sum n_i  \xi_i^{2/\alpha}}{\sum n_i  \xi_i^{1/\alpha}}% = \frac{M_0 }{\xi_0^{1/\alpha}} \xi_w
\end{eqnarray*}

%where $\xi_n$ and $\xi_w$ are defined respectively as the weight and number averaged "Frictional Dispersity Index", as 
%\begin{eqnarray*}
% \xi_n= \frac{\sum_i N_i\xi_i}{\sum_i N_i}
%\text{ and } 
%\xi_w = \frac{\sum_i N_i M_i\xi_i}{\sum_i N_i M_i}
%\end{eqnarray*}
For a given value of $\alpha$, we can compute from the distributions of Fig.~\ref{fig:Individual_Rubb}E the expected value for the ratio $M_\text{w}/M_\text{n}$. 

For $\alpha = 0.68$ \cite{wang2015scaling}, the extracted PDI would takes a value of $1.58 \pm0.17$.

For $\alpha = 1$, the extracted PDI would takes a value of $1.22 \pm0.05$.

A value $\alpha=1.68$ is necessary to match the experimental PDI of $1.07$.

These estimates suggest that the broad distribution of friction coefficient is set by intrinsic variations in chain conformation rather than sample polydispersity.

\subsection{Simulations} \label{sec:MM_simu}
To allow for direct comparison with our experimental results, our simulations model explicitly the global transport processes, as a combination of  through 3D near surface transport and local adsorption at the interface. Following explicit simulations of random trajectories, these distributions are then down-sampled at at a given time interval ($\Delta t=100$ ms for Fig.~\ref{fig:Phobe2}D-E).

\subsubsection{Hopping transport}

We describe below the practical implementation of the Brownian dynamics simulations for the computation of the distribution of hopping distance. The stochastic motion is accounted for by assuming that over discrete time steps $\delta t$ (taken typically between 1 to 1000 ns), the particle moves randomly with jumps drawn from a normal gaussian distribution of width $\sqrt{2D \delta t}$. The displacement increments in the x, y and z direction write:
\begin{equation}
\begin{split}
\delta x & = v(z)\delta t + \sqrt{2D \delta t} N(0,1) \\
   \delta y  & = \sqrt{2D \delta t} N(0,1) \\
   \delta z  & = \sqrt{2D \delta t} N(0,1)
   \end{split}
 \end{equation}
 Where the local velocity field  at the interface is taken as $v(z) =\dot \gamma (z+\lambda)$ with $\dot \gamma$ [s\textsuperscript{-1}]  the shear rate and $\lambda$ [m] is the slip length at the interface. $N(0,1)$ characterizes an elementary normal distribution.
%A major parameter in the simulations is the characteristic size of the elementary jumping length, expressed as $a = \sqrt{2D \delta t} $. This quantity if a function of the effective sampling time and diffusion coefficient. If we assume a diffusion coefficient $D=  4\cdot 10^{-11}$ m$^2$.s$^{-1}$, $a$ will be taking typical values of $0.894$, $2.83$, $8.94$ nm for $\delta t\in [10, 100, 1000]$ ns. 
Normalizing displacements by a characteristic distance $a = 1$ nm, we obtain in dimensionless form: \begin{equation}
 \begin{split}
    \delta \tilde x  &= \tilde \gamma (\tilde z + \tilde \lambda ) + \tilde \sigma  N(0,1) \\
     \delta \tilde y  &= \tilde \sigma  N(0,1) \\
 \delta \tilde z  &= \tilde \sigma  N(0,1) 
      \end{split}
 \end{equation}
 With $\tilde \sigma = \sqrt{2D\delta t}/a$, $\tilde \gamma = \dot \gamma \delta t$, $\tilde x=x/a$ and $\tilde z=z/a$.
 
 The Brownian particle starts from the interface ($z=0$) at time $t = 0$ and its motion is solved in the upper half space $z>0$.  As soon as the particle crosses the interface again ($z<0$) the simulation is stopped and the mean traveled distance $\Delta X$ and $\Delta Y$ is recorded. Only trajectories for which the initial step is away from the interface are considered. Representative trajectories in the presence of a biasing flow are shown in Fig. \ref{fig:Phile2}.

In order to probe the sanity of our simulations, we verified that our results are independent of the normalized simulation time, performing simulations for incremental time $\delta t = 10$, $100$  and $1000$ ns, and given a fixed diffusion coefficient $D = 4 \times 10^{-11}$  m\textsuperscript{2}.s\textsuperscript{-1}. These distributions can indeed be accurately collapsed over each other, except for the regime of small displacements $dx/dy$, which becomes dependent of $\tilde \sigma$.

%Jump distribution is described by a power-law -2.

%\begin{figure}[H]
%    \centering
%    \includegraphics[width=0.5\linewidth]{06_Simulations/06_Figures/ChapitreSimu-04.pdf}
%    \caption{\textbf{Evolution of hopping distributions at various shear rates}. XXXX. Simulation conditions: $\dot \gamma \in [0; 10^5]$ s\textsuperscript{-1}, $D = 4\cdot 10^{-11}$  m\textsuperscript{2}.s\textsuperscript{-1} and $\lambda =0$ nm. The power-law scaling at zero shear rate has power -2.}
%    \label{fig:simu_Arms}
%\end{figure}

%{\color{red} to do : run slip simulations with smaller dt and smaller gamamdot} why?...

 \subsubsection{Coupling with adsorption processes}

%\begin{figure}[H]
%    \centering
%  \includegraphics[width=0.8\linewidth]{A1_Simulations/A1_Figures/ShemaDiff-01.eps}
%  \caption{\textbf{Schematic of the scheme for  Brownian dynamics simulations}. We consider the alternance between 3D flights and periods where the molecule is physisorbed to the interface, in an adsorbed or freely 2D diffusing state.}
% \label{fig:simu_Traj_adsorption}
%\end{figure}

To model adsorption, we assumed a power-law scaling in the residence time (as observed experimentally), with a distribution $\psi(\tau)\sim \tau^{-\alpha}$, where $\alpha $ is the typically an exponent $\approx 2$, and a mean adsorption time $\tau_\text{m}$. In practice, the waiting time was derived from a uniform distribution $k\in [0,1]$ , as $\tau = \tau_\text{m} (1-k)^{1/(1-\alpha)}$.

\bibliography{references}% Produces the bibliography via BibTeX.

\newpage
\appendix
\section{Supplementary Information}

\counterwithin{figure}{section} % Optional: If you want figures numbered as "SI1", "SI2", etc.
\renewcommand{\thefigure}{SI.\arabic{figure}} % Redefine figure numbering format
\setcounter{figure}{0} % Reset figure counter to 0

\begin{figure}[htb!]
    \centering
\includegraphics[width=1\linewidth]{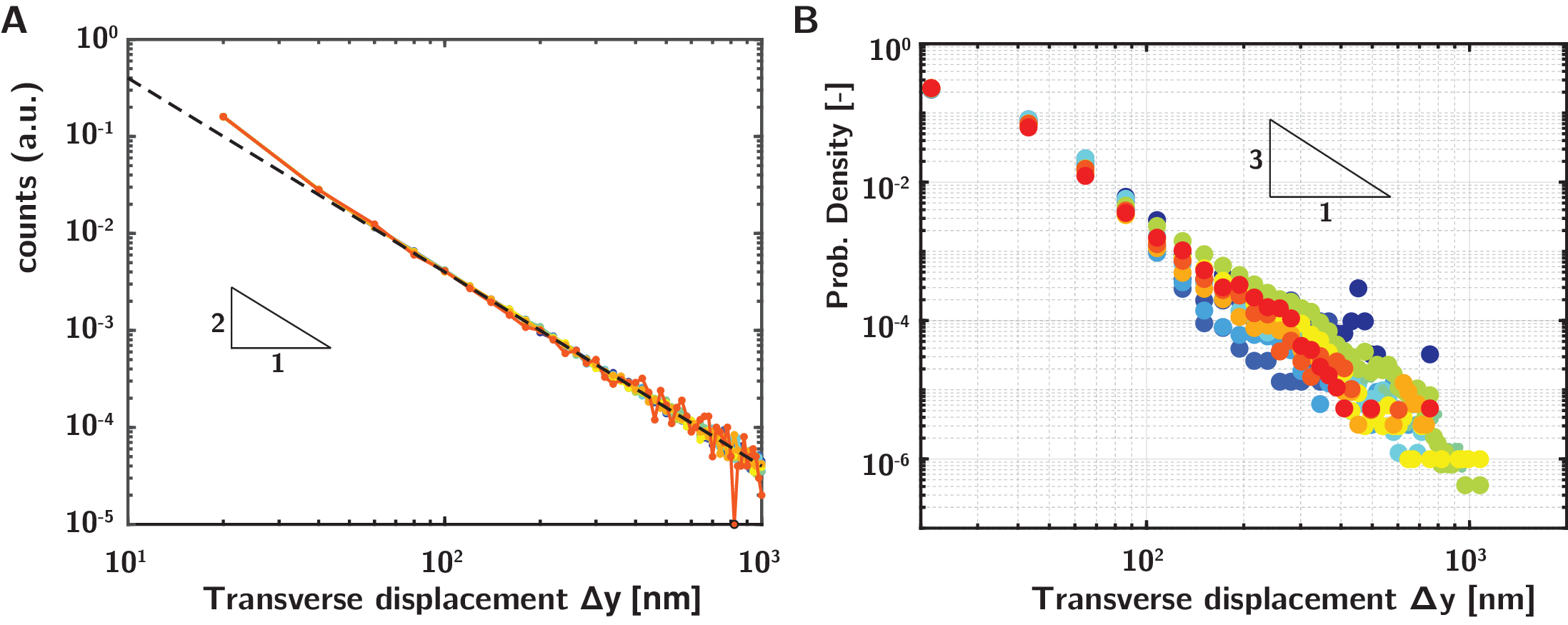}
    \caption{ \textbf{Comparison long-distance transport scaling.} \textbf{(A)} Computed transverse (equilibrium) displacement distributions.  \textbf{(B)} Experimentally observed displacement distributions (same color code as in Fig. \ref{fig:Phile2}).}
    \label{fig:simu_LongDistance}
\end{figure}

\begin{figure}[htb!]
    \centering
\includegraphics[width=0.8\linewidth]{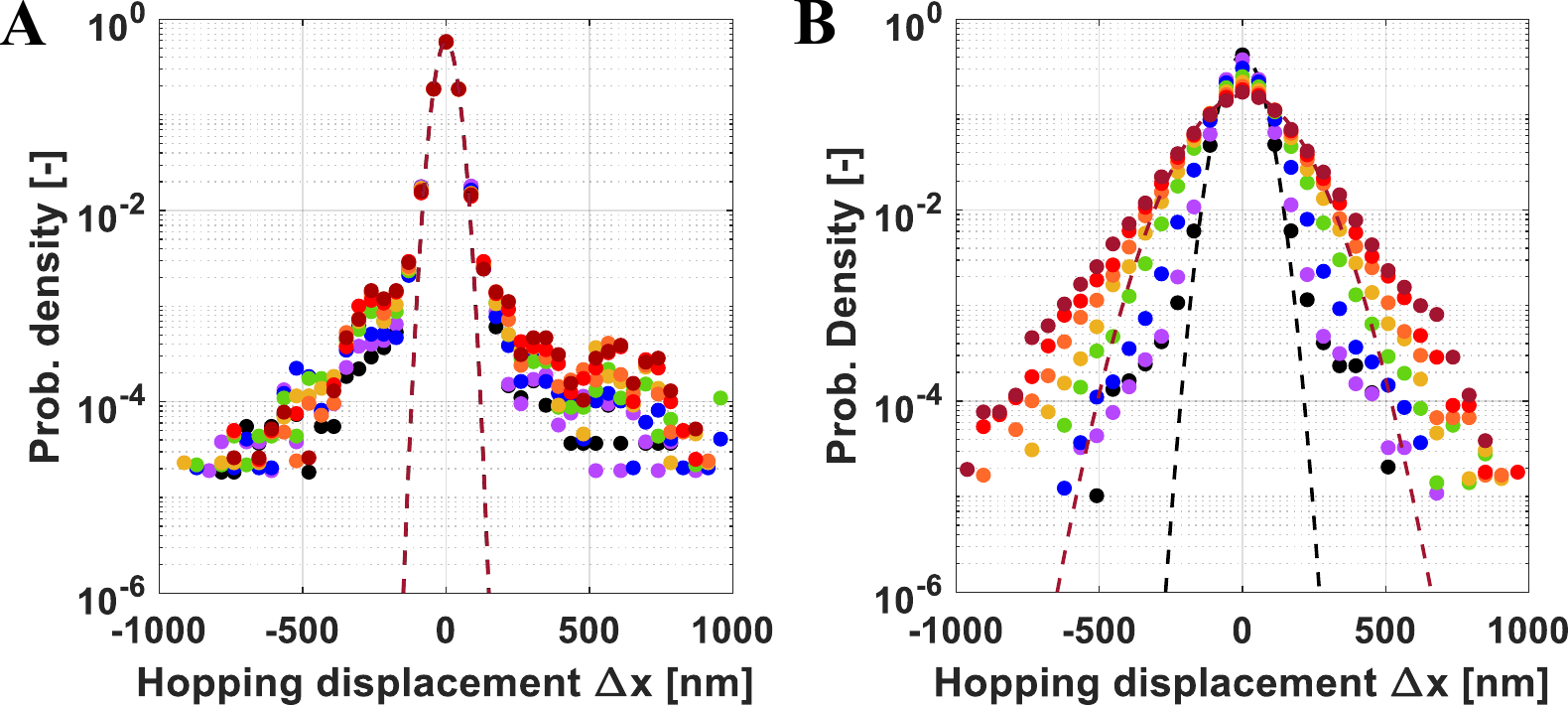}
    \caption{\textbf{Displacement distribution}. \textbf{(A)} and \textbf{(B)} Distributions of the elementary displacements of the macromolecules on bare glass (A) and OTS (B) surfaces. Experimental time $\Delta$t is artificially increased from 12 (black) to 360 (dark red) ms. For lowest and highest $\Delta$t, a Gaussian fit of the central peak of the distribution is shown.}
    \label{fig:DistribEq}
\end{figure}

\begin{figure}[htb!]
    \centering
\includegraphics[width=\linewidth]{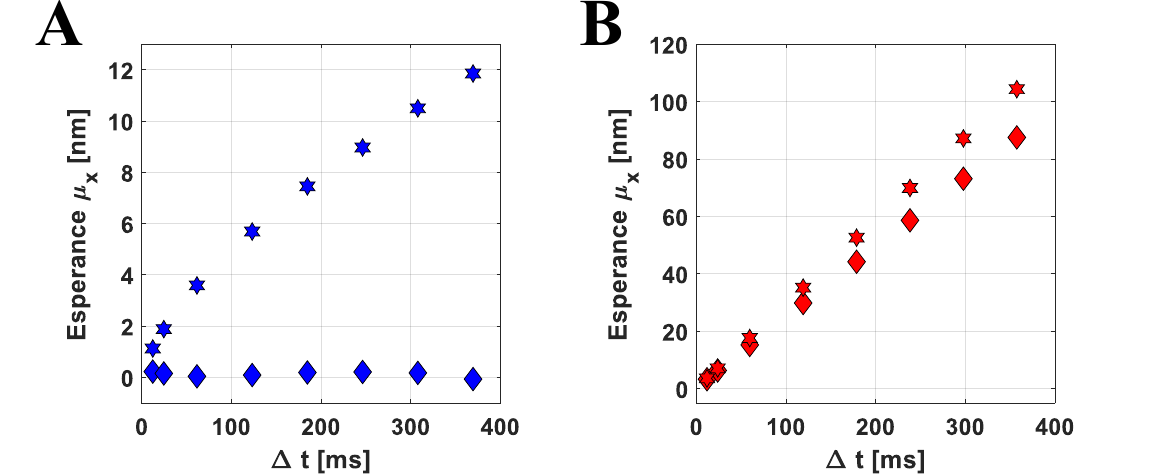}
    \caption{\textbf{(A)} Esperance on hydrophilic surface extracted from the Gaussian fit (diamonds) or calculated from the whole distribution (stars) parallel to the flow.
    \textbf{(B)} Esperance on hydrophobic surface extracted from the Gaussian fit (diamonds) or calculated from the whole distribution (stars) parallel to the flow . }
    \label{fig:SI_mu}
\end{figure}

\end{document}